\documentclass[preprint,11pt]{elsarticle}

\usepackage{lineno}
\usepackage{hyperref}
\usepackage[]{units}
\usepackage{subcaption}
\usepackage{amsmath}
\usepackage{xspace}
\usepackage{marvosym}
\usepackage{color}
\usepackage{amsmath,amssymb} 

\begin{document}

\begin{frontmatter}

\title{Astrophysical interpretation of small-scale neutrino angular correlation searches with IceCube}

\author[rwth]{Martin Leuermann} 
\author[rwth,buw]{Michael Schimp\corref{cor1}}
\author[rwth]{Christopher H. Wiebusch}
\address[rwth]{III. Physikalisches Institut B, RWTH Aachen University, Aachen, Germany}
\address[buw]{Now at Bergische Universit\"at Wuppertal, Fachbereich C -- Physik, Wuppertal, Germany}

\cortext[cor1]{Corresponding author (\href{mailto:michael.schimp@rwth-aachen.de}{\texttt{michael.schimp@rwth-aachen.de}})}

\date{Version of \today}

\journal{Astroparticle Physics}





\begin{abstract}
The IceCube Neutrino Observatory has  discovered a diffuse all-flavor flux of high-energy  astrophysical neutrinos.
However, the corresponding astrophysical sources have not yet been identified.
Neither significant point sources 
nor significant
angular correlations of event directions 
have been observed by IceCube or other instruments to date.
We present a new method to interpret the non-observation
of angular correlations in terms of exclusions on the strength and number of 
point-like neutrino sources in generic astrophysical scenarios.
Additionally, we constrain the presence of these sources taking into account
the measurement of the diffuse high-energy neutrino flux by IceCube.
We apply the method to two types of astrophysically motivated 
source count distributions:
The first type is obtained by considering the cosmological evolution of the co-moving density 
of active galaxies, while the second type is directly derived 
from the gamma ray source count distribution observed by  Fermi-LAT. 
As a result, we constrain the possible parameter space for both types of source count distributions.

\end{abstract}

\end{frontmatter}

\section{Introduction\label{sec:intro}}

\subsection{Astrophysical neutrino observation by IceCube}

The IceCube Neutrino Observatory \cite{IC} at the Geographic South Pole 
has discovered an all-sky 
diffuse flux
of high-energy cosmic neutrinos \cite{Aartsen:2014gkd,Aartsen:2013jdh}
based on neutrinos of all flavors interacting within the detector.
However, no astrophysical sources of this flux could be identified yet.
Recently, this all-flavor flux has been confirmed by the measurement of an 
excess of uncontained up-going muons \cite{Aartsen:2015rwa} at high energies
above the background originating from interactions of atmospheric neutrinos.
These muons are produced by charged current interactions of muon neutrinos
in the ice, where the direction of the muon and the neutrino agree well within $\sim 1^{\circ}$ in the considered energy range.
Muons propagate large distances through the ice, and 
can be measured with  good angular resolution, i.e.  $<1^{\circ}$.
Though such events are ideally suited for the identification of the sources,
neither searches for angular autocorrelations of neutrino arrival directions nor correlations of neutrino arrival directions
with the positions of known astrophysical sources have resulted
in a significant observation \cite{Aartsen:2014cva,Aartsen:2014ivk}.
In conclusion, the total number of sources of the observed flux is presumably
large as so far the individual sources have been  too weak to be 
detectable with respect to the atmospheric neutrino background.

\subsection{Angular correlations of neutrino arrival directions}

This paper focusses on the non-observation of an angular correlation within
108\,310  up-going muons  in  
 IceCube data measured from 2008 to 2011 with the detector
configurations  \emph{IC40}, \emph{IC59} and \emph{IC79} \cite{Aartsen:2014ivk}.
That result was obtained based on two analyses. The first is 
 a binned correlation analysis and the second uses the 
power coefficients of a multipole expansion of the sky map 
of detected neutrino arrival directions.
In this work, we focus on the second result.
Here, weak sources, constituting the signal, were assumed to be 
isotropically distributed over the sky. The signal was 
benchmarked according to different signal hypotheses, characterized by three quantities:
the total \emph{number of sources} in the full sky $N_{\mathrm{Sou}}$, a 
universal \emph{strength} of each source $\mu $, and the \emph{spectral index}
$\gamma$ of the energy spectrum.
The parameter $\mu$ is  the mean number of measured neutrinos per
 source at the horizon.
While at the horizon the detection efficiency is largest, each source off the horizon is assigned a lower number of neutrinos according to the declination dependent detector acceptance.

\begin{figure}
\includegraphics[width=\linewidth]{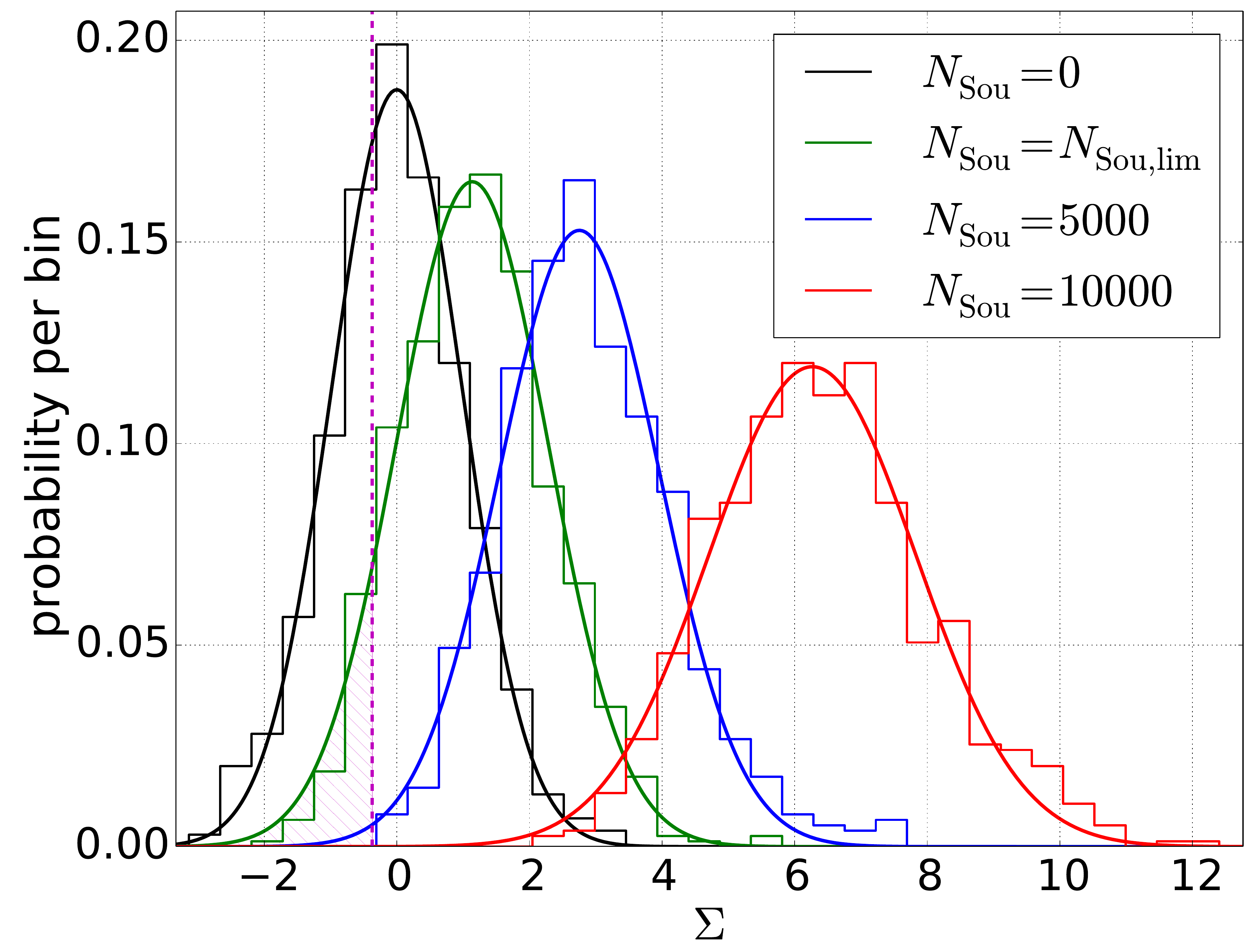}
\caption{Distributions of the test statistic (TS) for different numbers of sources in the full sky $N_\mathrm{Sou}$, fixed source strength $\mu=3$ and energy spectrum $\gamma=2.5$; dashed vertical line: result from the 
experimental skymap $\Sigma_{\mathrm{exp}} = -0.3$  \cite{Aartsen:2014ivk}; 
hatched area: lower 10\% quantile of the TS distribution for a signalness $\Sigma_{\mathrm{lim}} = 1.07$ corresponding to observed the upper limit.}
\label{fig:TSN}
\end{figure}

\begin{figure}
\includegraphics[width=\linewidth]{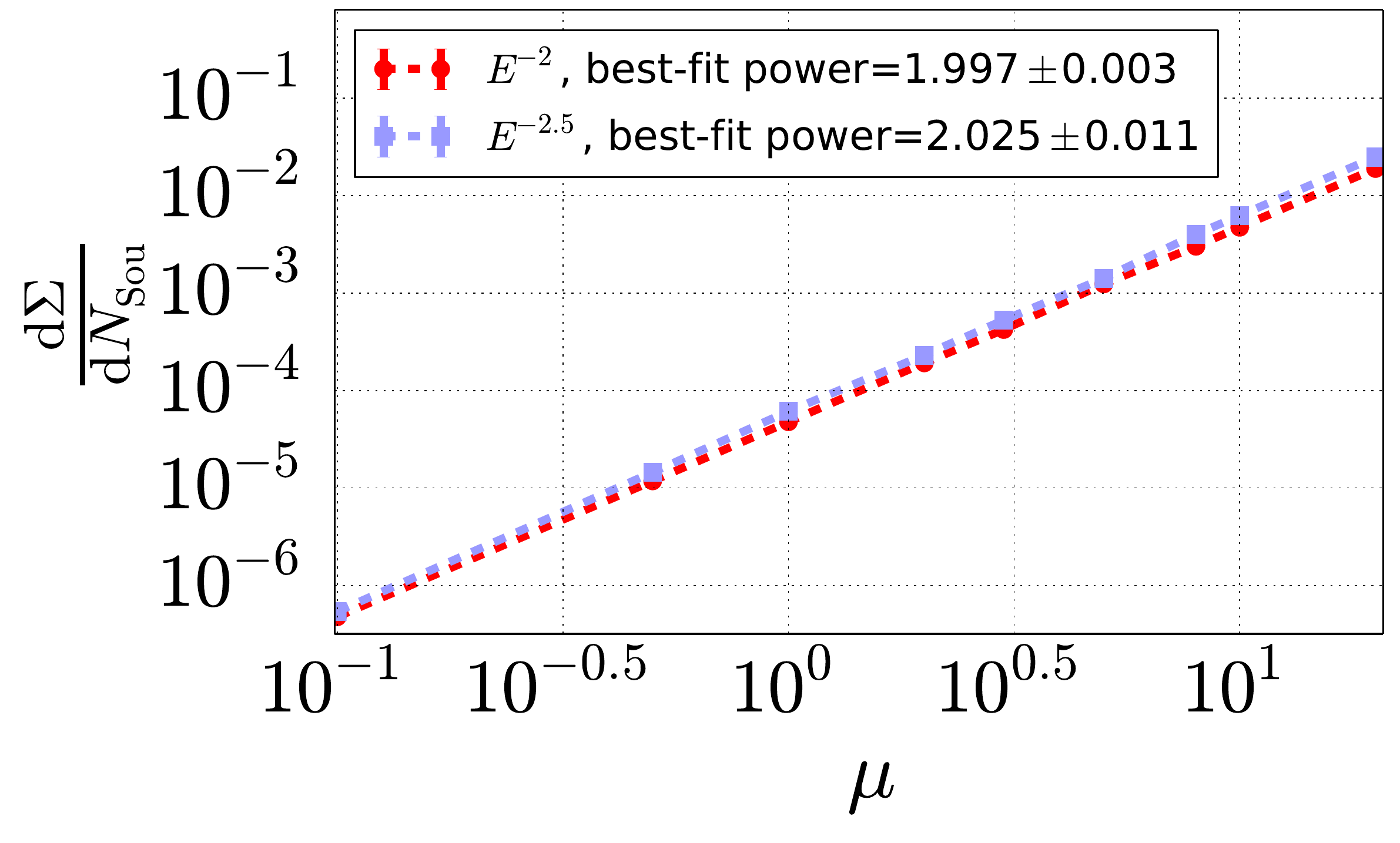}
\caption{Signalness per source $\frac{\mathrm{d}\Sigma}{\mathrm{d}N_{\mathrm{Sou}}}$ against source strength $\mu$ for astrophysical energy spectra $E^{-2}$ and $E^{-2.5}$; legend: exponent of best-fit power law.}
\label{fig:signalness_per_source}
\end{figure}

The analysis from \cite{Aartsen:2014ivk}
uses a test statistics~(TS) 
that denotes how significantly the angular correlations of muon directions in the specific skymap are distinguishable from the random atmospheric background.
The expected TS shift for signal with respect to the TS expectation
for pure atmospheric background  in units of the standard deviation
 of the background TS is called \emph{\textbf{signalness}} $\Sigma$ in the following.
In Figure~\ref{fig:TSN}, the TS distributions for signal hypotheses with different values for $N_\mathrm{Sou}$ are shown. 
The distributions are obtained by simulations of random skymaps 
using the information from \cite{Aartsen:2014ivk} about the point spread 
function and zenith-dependent detector acceptance.
We find that for a fixed source strength the signalness, i.e. the mean of the TS distribution, scales~$ \propto N_\mathrm{Sou} $.
In Figure~\ref{fig:signalness_per_source}, the signalness per source is shown
as a function of the source strength~$\mu $. 
We find that the signalness per source increases with stronger sources consistently with  $\frac{\mathrm{d}\Sigma}{\mathrm{d}N_{\mathrm{Sou}}} \propto \mu^2 $, independent  of the assumed  energy spectrum.


The experimentally observed value is $\Sigma_{\mathrm{exp}} = -0.3$ \cite{Aartsen:2014ivk}. The corresponding exclusion limits on the number of sources 
  $N_\mathrm{Sou}=N_\mathrm{Sou,lim}$ for different values of $\mu $  are obtained from simulations as those values of $ N_\mathrm{Sou} $ for which 90\%  of 
experiments would result in a larger signalness than the observed $\Sigma_{\mathrm{exp}} $.  
We find that for all different combinations of $ N_\mathrm{Sou} $ and $\mu $ this results in the same signalness $\Sigma $, 
while the variance of the TS distribution is identical.
Correspondingly,
the median  signalness corresponding to the observed upper limit is 
$\Sigma_{\mathrm{lim}} = 1.07$ and 
does not depend on the specific choice of signal parameters.
Thus, $N_\mathrm{Sou,lim}$ is expressed as a simple function of the source strength $\mu $. 

\subsection{Purpose of this work}

Purpose of this work is to re-interpret the given exclusion limits for the number of sources of a fixed source strength in terms  of astrophysically motivated distributions
of source strengths~$\frac{\mathrm{d}N_{\mathrm{Sou}}}{\mathrm{d}\mu}$.
To do this, we calculate the expected signalness as a function of the respective astrophysical model parameters and compare this to the experimentally 
excluded signalness.
For this, we make use of the dependencies of the signalness on the model parameters  $N_\mathrm{Sou}$ and $\mu $ as introduced above.
As benchmark scenarios, we use
two astrophysical models.
For the first model, we assume isotropically distributed sources with a number density 
following the red-shift dependent evolution of active galactic nuclei (AGNs).
For the second model, we assume an isotropic distribution of sources with strengths
analogous to the strengths of extragalactic sources of high-energy photons as observed by the Fermi-LAT satellite.
Further details of the models that were taken into
account in this work are given in Section~\ref{subsec:scd_calc}.

Additionally, we take into account the measured diffuse astrophysical
muon neutrino flux from the Northern hemisphere \cite{Aartsen:2015rwa} in oder to further constrain the scenarios.
For both of the mentioned models, we test two 
astrophysical neutrino energy spectra $\propto E^{-\gamma}$ that 
are compatible with this measurement.
That is a hard spectral index of $\gamma=2.0$ and a soft spectral index of $\gamma=2.5$.

It should be noted that other interpretations of a diffuse astrophysical neutrino flux---before and after the measurement by IceCube---have been published.
These include different approaches as, for example, the interpretation of
diffuse and/or stacking limits in terms of different production mechanisms
\cite{becker2007astrophysical} or the presence of point sources and their neutrino power density \cite{silvestri2010constraints}.
One recent approach is to constrain the presence of sources that are obscured in gamma rays but well visible in neutrinos such as choked GRBs \cite{murase2015hidden}.
Our approach differs from these in the manner that we additionally (and, in fact, primarily) interpret the absence
of angular correlations in neutrino directions rather than the diffuse astrophysical flux.
Taking this flux into account to further constrain our paramters of interest
is technically just an optional addition but is still meaningful due to the relevance of this flux measurement.
Also, while we apply our approach to specific source scenarios in this work, the method we present is generally applicable for other scenarios.

\section{Method}

\subsection{Calculation of the source count distributions}\label{subsec:scd_calc}

\subsubsection{Cosmologically distributed sources}\label{subsubsec:cosmo_calc}

For the application to sources motivated by the cosmological evolution of AGN,
we assume standard sources that  exhibit  the same  muon neutrino luminosity 
 $L$ in the energy range from  $\unit[100]{GeV} $ to $ \unit[100]{TeV}$ used in the IceCube angular correlation analysis.
Due to red-shift of energy this  leads to a red-shift dependency of the
energy range that is used for the luminosity normalization.
Using $L$, the source strength $\mu$ can be expressed in dependence on the cosmological redshift $z$:
\begin{align}
\label{eqn:mu(z)}\mu(z) & = \frac{L}{4\pi d_L^2(z)\cdot (1+z)^{\gamma-2}}\cdot b(\gamma)
\end{align}
where $d_\mathrm{L}(z)$ is the luminosity distance.
The factor 
\begin{align}
\label{eqn:b(gamma)}b(\gamma) & = \frac{\sum\limits_{\mathrm{IC}}T^{\mathrm{IC}}\int\limits_{\unit[100]{GeV}}^{\unit[100]{TeV}}\mathrm{d}E\, A_{\mathrm{eff}}^{\mathrm{IC}}(E)E^{-\gamma}}{f(\gamma)\cdot\int\limits_{\unit[100]{GeV}}^{\unit[100]{TeV}}\mathrm{d}E\, E^{1-\gamma}}
\end{align}
takes into account the observational parameters where  $T^{\mathrm{IC}}$ denotes the livetime of IceCube for the operation of each detector configuration $\mathrm{IC}$ and $A_{\mathrm{eff}}^{\mathrm{IC}}(E)$ is the declination-averaged effective area of each configuration  $\mathrm{IC}$.
The factor $f(\gamma) $ is the declination-averaged detector
acceptance divided by the detector acceptance at the horizon.
It compensates the usage of the declination-averaged effective area $A_{\mathrm{eff}}^{\mathrm{IC}}$ in order to obtain the expected number of neutrinos per source at the horizon $\mu$ instead of a declination-averaged expected number of neutrinos per source.
The values for $f(\gamma)$ are $0.624$ and $0.848$ for energy spectra with $\gamma=2.0$ and $\gamma=2.5$, respectively. 

\begin{figure}[htp]
\includegraphics[width=\textwidth]{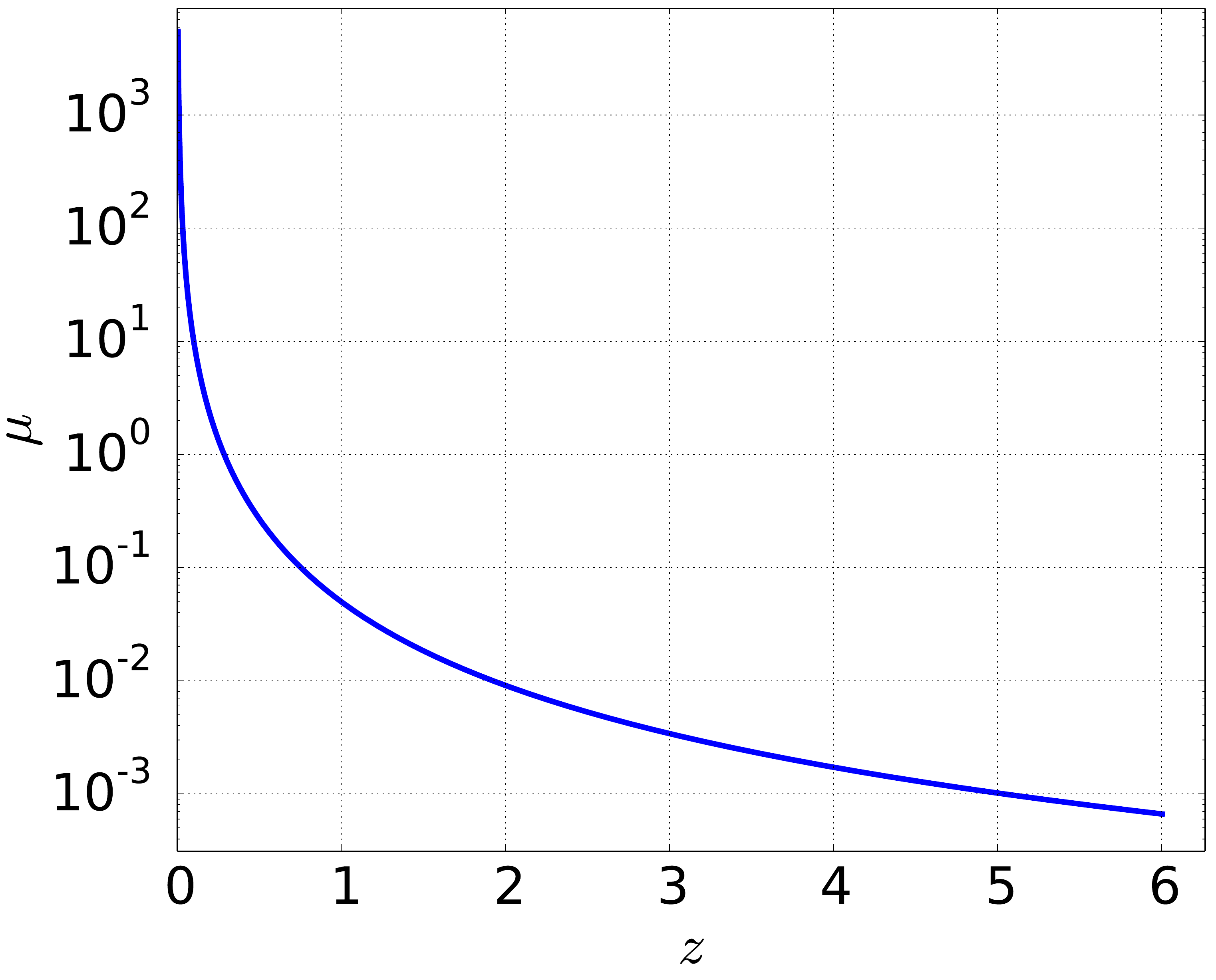}\caption{Redshift-dependent source strength $\mu(z)$ for sources with universal muon neutrino source luminosities $L = \unit[7\cdot10^{44}]{\frac{erg}{s}}$ and energy spectra with $\gamma=2.0$}\label{fig:scratch_mu_z}
\end{figure}

The luminosity distance $d_\mathrm{L}(z)$ and the co-moving volume 
$V_\mathrm{c}(z)$, are calculated using the cosmology calculator described in \cite{Wright}. 
For this, the following cosmological parameter values are  assumed \cite{PDG}:
 $\Omega_{\mathrm{m}}=0.315^{+0.016}_{-0.017}$, $\Omega_\mathrm{r}=8.53\cdot 10^{-5}$ 
and  $\Omega_{\mathrm{\Lambda}}=0.685^{+0.017}_{-0.016}$.
The given errors are propagated for a cross check:
The resulting relative errors for the total number of measured neutrinos $n(\mathrm{scd})$ and the signalness $\Sigma$ are $\sim2.3\%$ and $1.8\%$, respectively.
They are not regarded further as they have very little impact on the results (see for comparison the scale of $\alpha$ in Figure~\ref{fig:cosmo}).
In Figure~\ref{fig:scratch_mu_z}, an exemplary distribution for $\mu(z)$ is given.

\begin{figure}
\includegraphics[width=\textwidth]{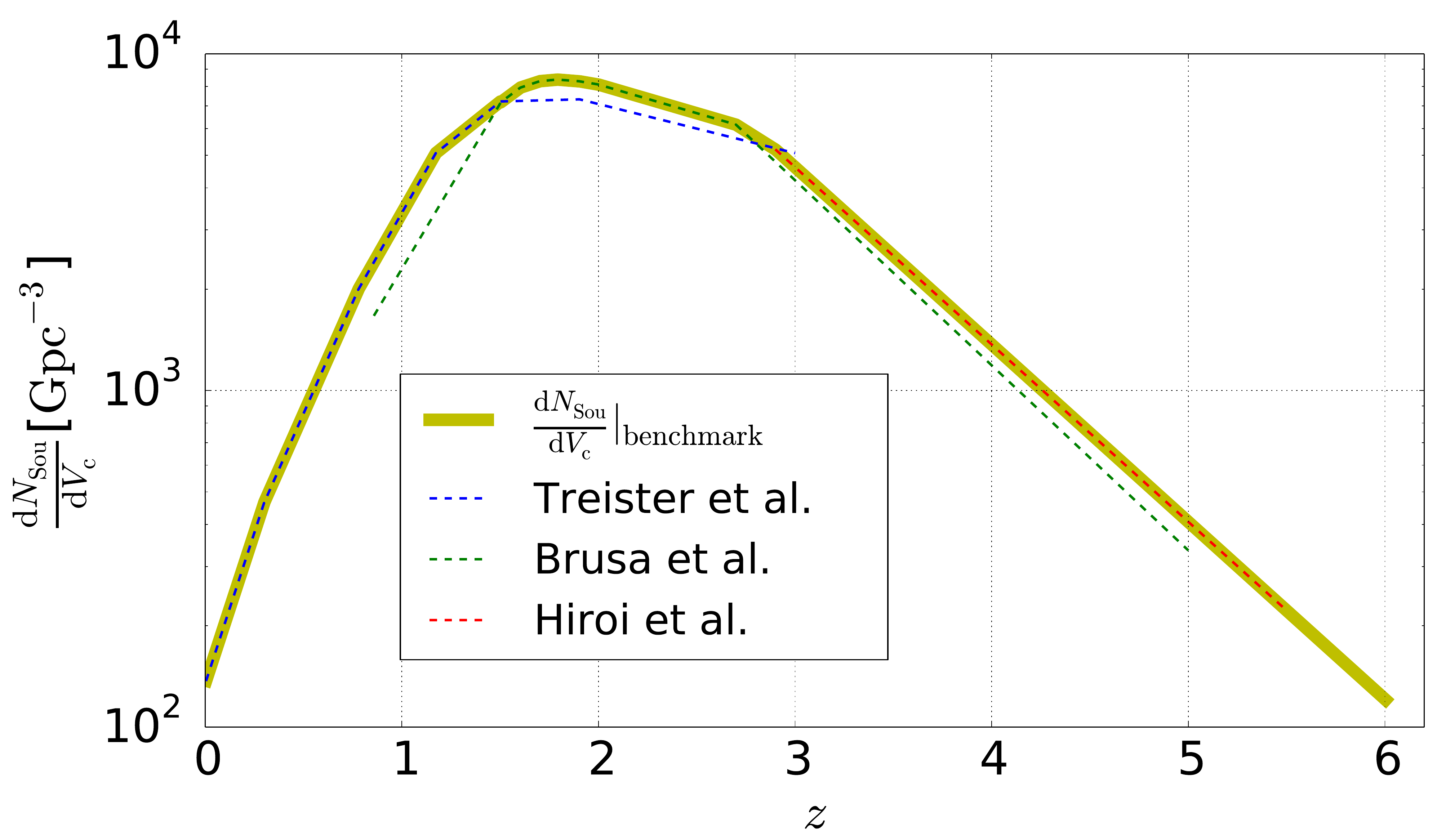}\caption{Yellow solid line: combined redshift-dependent co-moving source density representing AGN densities up to $z=6$ \cite{treister,brusa,hiroi}, i.e. $\frac{\mathrm{d}N_\mathrm{Sou}}{\mathrm{d}V_\mathrm{c}}\big|_\mathrm{benchmark}$; dashed lines: contributions from the articles given in the legend; kinks originate from the provision of the density distributions as interval-wise power laws}\label{fig:source_dens}
\end{figure}

To account for the evolution of sources, a redshift-dependent  co-moving source density $\frac{\mathrm{d}N_\mathrm{Sou}}{\mathrm{d}V_\mathrm{c}}(z)$ has to be assumed.
As a benchmark the distribution  $\frac{\mathrm{d}N_\mathrm{Sou}}{\mathrm{d}V_\mathrm{c}}\big|_\mathrm{benchmark}$ is constructed  by combining the distributions given in: \cite{treister}, \cite{brusa} and \cite{hiroi}.
These individual distributions are shown as dashed lines in Figure~\ref{fig:source_dens}.
They are fits to models, representing the redshift-dependent co-moving density of high luminosity AGN above X-ray  luminosity thresholds of $\sim\unit[10^{44}]{\frac{erg}{s}}$ up to high redshifts.
The resulting distribution is represented by the yellow wide line in Figure~\ref{fig:source_dens}.
The redshift of the closest known AGN in the Northern Sky~(M87) \cite{M87} is $0.004$, while the expected contributions to the signalness~$\Sigma$ and to the number of measured neutrinos $\mu(z)$ are negligible for $z>6$.
Thus, $\frac{\mathrm{d}N_\mathrm{Sou}}{\mathrm{d}V_\mathrm{c}}\big|_\mathrm{benchmark}$ is set to zero for $z<0.004$ and $z>6$.
For this  work, only the shape of $\frac{\mathrm{d}N_\mathrm{Sou}}{\mathrm{d}V_\mathrm{c}}\big|_\mathrm{benchmark}$ is relevant, because  the absolute scale is considered as a free parameter in our calculations (see below).
Therefore, the uncertainties on the absolute values of 
$\frac{\mathrm{d}N_\mathrm{Sou}}{\mathrm{d}V_\mathrm{c}}\big|_\mathrm{benchmark}$ are not taken into account in the following.

\begin{figure}[htp]
\includegraphics[width=\textwidth]{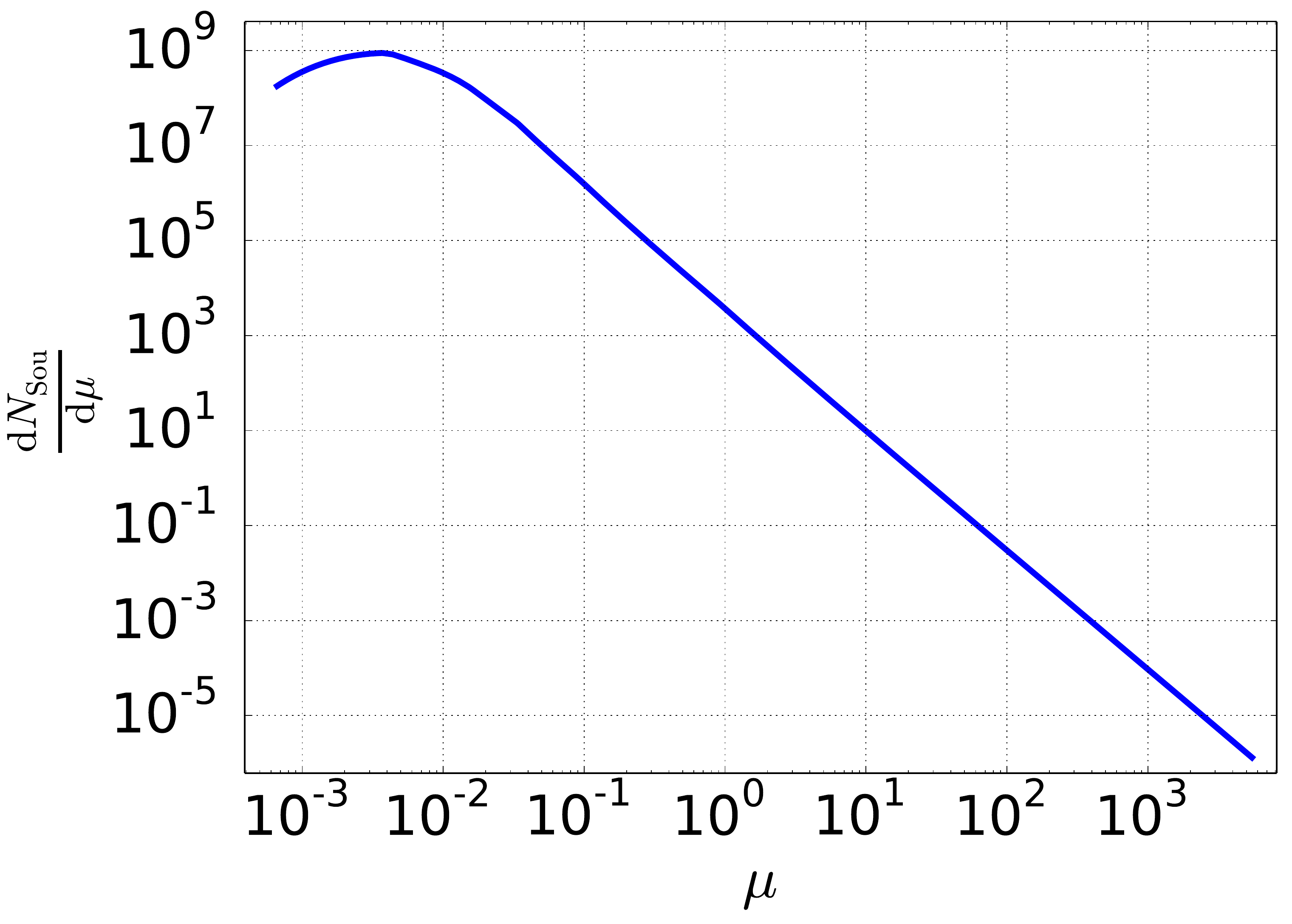}\caption{Source count distribution for the benchmark redshift-dependent co-moving source density distribution $\frac{\mathrm{d}N_\mathrm{Sou}}{\mathrm{d}V_\mathrm{c}}\big|_\mathrm{benchmark}$ (i.e. $\alpha = 1$), $L=\unit[7\cdot10^{44}]{\frac{erg}{s}}$ and $\gamma=2.0$}\label{fig:scratch_var_scd}
\end{figure}

Given $\mu(z)$ and the co-moving source density $\frac{\mathrm{d}N_\mathrm{Sou}}{\mathrm{d}V_\mathrm{c}}$, the source count distribution  is calculated by:
\begin{equation}\label{eqn:scratch_scd}\frac{\mathrm{d}N_{\mathrm{Sou}}}{\mathrm{d}\mu}(\mu)=-\frac{\mathrm{d}z}{\mathrm{d}\mu}\frac{\mathrm{d}N_\mathrm{Sou}}{\mathrm{d}V_\mathrm{c}}\frac{\mathrm{d}V_\mathrm{c}}{\mathrm{d}z}\end{equation}
In Figure~\ref{fig:scratch_var_scd}, the resulting  source count distribution 
is shown for our benchmark model  $\frac{\mathrm{d}N_\mathrm{Sou}}{\mathrm{d}V_\mathrm{c}}\big|_\mathrm{benchmark}$.

Besides the universal muon neutrino luminosity $L$, 
a scale factor $\alpha$ for the source density is used as a second model parameter
\begin{equation}\label{eqn:alpha}
\frac{\mathrm{d}N_\mathrm{Sou}}{\mathrm{d}V_\mathrm{c}}=\alpha\frac{\mathrm{d}N_\mathrm{Sou}}{\mathrm{d}V_\mathrm{c}}\big|_\mathrm{benchmark}
\end{equation}
Thus, constraining or predicting a certain value of $\alpha$ is equivalent 
to constraining or predicting the normalization of the source density 
distribution. 
As a consequence, $\alpha$ can be interpreted as a relative 
fraction of AGN described by the benchmark source 
density 
$\frac{\mathrm{d}N_\mathrm{Sou}}{\mathrm{d}V_\mathrm{c}}\big|_\mathrm{benchmark}$
which contribute to the observed signal.

\subsubsection{Fermi-LAT extragalactic sources}\label{subsubsec:Fermi}

\begin{figure}[htp]
\includegraphics[width=\linewidth]{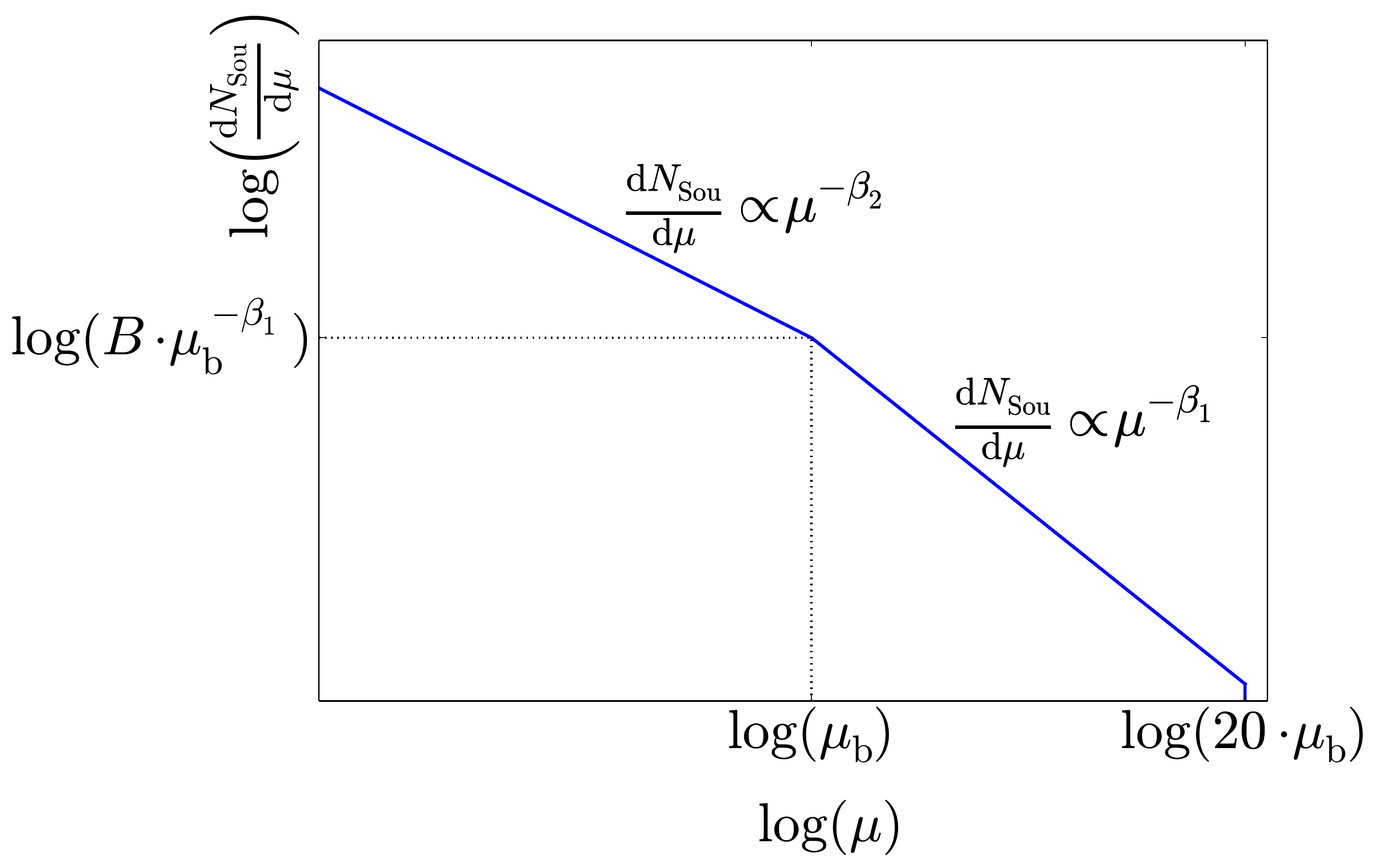}
\caption{Sketch of the source count distribution $\frac{\mathrm{d}N_{\mathrm{Sou}}}{\mathrm{d}\mu}(\mu)$ with powers $\beta_1$ and $\beta_2$ adopted from Fermi-LAT}
\label{fig:scd}\end{figure}

The gamma ray telescope Fermi-LAT has measured the photon flux of extragalactic high-energy photon sources with a fitted average energy spectrum of $E^{-2.4}$ in the energy range from $\unit[100]{MeV}$ to $\unit[100]{GeV}$ in a high-latitude survey \cite{Fermi}.
It is parametrized by a broken power law:
\begin{equation}\label{eqn:scd_original}
\frac{\mathrm{d}N_{\mathrm{Sou}}}{\mathrm{d}S}(S) = \left\{\begin{array}{r r}AS^{-\beta_1}, & S\geq S_{\mathrm{b}} \\ AS_{\mathrm{b}}^{-\beta_1+\beta_2}S^{-\beta_2}, & S<S_{\mathrm{b}}\end{array}\right.,
\end{equation}
where $\beta_1 = 2.49\pm0.12$ and $\beta_2 = 1.58\pm0.08$ are the powers of the source count distribution after and before the break, respectively.
$A = \unit[(16.46\pm0.80)\cdot10^{-14}]{cm^2\,deg^{-2}\,s}$ is a normalization
 factor for $\frac{\mathrm{d}N_{\mathrm{Sou}}}{\mathrm{d}S}$ and $S_{\mathrm{b}} = \unit[(6.60\pm0.91)\cdot10^{-8}]{cm^{-2}\,s^{-1}}$ is the photon flux at the break of the source count distribution.
In Figure~\ref{fig:scd}, an illustration of the used parametrization is shown.

As these photon sources are also candidates for high-energy neutrino sources \cite{Elisa}, we adopt 
the parametrization
as a neutrino source count distribution as explained in the following.

The photon fluxes  $S$ and $S_{\mathrm{b}}$ are replaced by  
the neutrino source strength $\mu$ as defined in Section~\ref{sec:intro}
and the source strength at the break is denoted $\mu_{\mathrm{b}}$.
The normalization $A$ is replaced by the dimensionless factor $B$.
Hence, all neutrino source count distribution parameters are 
dimensionless in contrast to the parameters measured by Fermi-LAT.
Furthermore, a cutoff is introduced by setting the source count distribution to zero for all source strengths above 
a maximum $\mu_{\mathrm{max}}$ to avoid 
divergences in the signalness calculation.
The cutoff  is fixed to $\mu_{\mathrm{max}} = 20\mu_{\mathrm{b}}$ corresponding to the brightest source in the Fermi-LAT sample which has a flux of $S_{\mathrm{max}}\approx 20 S_{\mathrm{b}}$ \cite{Fermi}.
This results in the following parametrization for the neutrino source count distribution:
\begin{equation}\label{eqn:scd}
\frac{\mathrm{d}N_{\mathrm{Sou}}}{\mathrm{d}\mu}(\mu) = \left\{\begin{array}{r r} 0, & \mu\geq20\mu_{\mathrm{b}}\\ \hfill B\mu^{-\beta_1}, & 20\mu_{\mathrm{b}}>\mu\geq\mu_{\mathrm{b}} \\ B\mu_{\mathrm{b}}^{-\beta_1+\beta_2}\mu^{-\beta_2}, &  \mu<\mu_{\mathrm{b}}\end{array}\right.
\end{equation} 
The best-fit values for the powers of the Fermi-LAT source count distribution, $\beta_1 = 2.49$ and $\beta_2 = 1.58$, are initially 
also applied for the neutrino source count distributions before we study more general values in 
Section~\ref{subsec:varybeta}.

Though motivated by the Fermi-LAT observations, there is no a-priori correct conversion between the parameters describing the neutrino and photon source distributions, $(\mu_\mathrm{b},B)$ and $(S_\mathrm{b},A)$.
In particular, 
the sensitive energy ranges for the Fermi-LAT high-latitude survey, $\unit[100] {MeV} \text{--} \unit[100] {GeV}$,  and for the IceCube measurement, $\unit[100] {GeV} \text{--} \unit[100] {TeV}$, differ.
However, as a benchmark we assume a universal neutrino-to-photon ratio $\varepsilon_{\nu/\gamma}$
for the flux received from these sources.
This ratio is assumed to be constant for all energies and for all sources of the given population.
One should note that several processes at the sources like inverse Compton scattering, bremsstrahlung and proton-synchrotron radiation might intorduce a bias to this ratio because they affect the correlation of photon and neutrino production in an energy dependent way \cite{zacharias2012external,eichmann2012differences,mucke2001proton}.

Using our assumption of a universal $\varepsilon_{\nu/\gamma}$, different values of $\varepsilon_{\nu/\gamma}$ scale the neutrino 
flux per source by the same factor for each source. Thus, they also scale the source strength $\mu$ of each source by the same factor  $\varepsilon_{\nu/\gamma}$
and the source count distribution can still be parametrized by the broken power law given in Equation~\eqref{eqn:scd}.

To relate the universal neutrino-to-photon ratio to the source count distribution parametrization from Fermi-LAT, first the values of $(\mu_{\mathrm{b}},B)$ for $\varepsilon_{\nu/\gamma} = 1$
are determined which are called $(\mu_{\mathrm{b,Fermi}},B_{\mathrm{Fermi}})$ in the following.
They are calculated by:
\begin{align}
\label{eqn:mubfermi}\mu_\mathrm{b,Fermi}&=a(\gamma)\cdot1000^{1-\gamma}\cdot S_{\mathrm{b}}\\
\label{eqn:Bfermi}B_\mathrm{Fermi}&=\left(a(\gamma)\cdot1000^{1-\gamma}\right)^{\beta_1-1}\cdot A\\
a(\gamma) & = \frac{\sum\limits_{\mathrm{IC}}T^{\mathrm{IC}}\int\limits_{\unit[100]{GeV}}^{\unit[100]{TeV}}\mathrm{d}E\, A_{\mathrm{eff}}^{\mathrm{IC}}E^{-\gamma}}{f(\gamma)\int\limits_{\unit[100]{GeV}}^{\unit[100]{TeV}}\mathrm{d}E\,E^{-\gamma}},
\end{align}
where $a(\gamma)$ is the factor converting a particle flux into the observed source strength $\mu$.
The factors of $1000^{1-\gamma}$ in Equations~\eqref{eqn:mubfermi} and~\eqref{eqn:Bfermi} take into account the different energy ranges of Fermi-LAT and IceCube for the assumed energy spectra.

For each value of $\varepsilon_{\nu/\gamma}\neq1$, the source strength $\mu$ of each source in the population with $B = B_{\mathrm{Fermi}}$ and $\mu_{\mathrm{b}} = \mu_{\mathrm{b,Fermi}}$ has to be multiplied by $\varepsilon_{\nu/\gamma}$.
For the source count distribution parameters $B$ and $\mu_{\mathrm{b}}$, this leads to: 
\begin{equation}\label{eqn:enugamma}
B=\varepsilon_{\nu/\gamma}^{\beta_1-1}\cdot B_{\mathrm{Fermi}}
\qquad \mbox{and} \qquad
\mu_{\mathrm{b}}=\varepsilon_{\nu/\gamma}\cdot\mu_{\mathrm{b,Fermi}}
\end{equation}

Note that we do not explicitly assume any absorption effects for the photons observed by Fermi-LAT with respect to the neutrinos observed by IceCube.
However, this issue is implicitly addressed in Section~\ref{subsec:varybeta} where variations in $\beta_1$ and $\beta_2$ can partly account for corresponding effects.

\subsection{Limit conversion}\label{subsec:conversion}

For the interpretation  of the limits from the angular correlation analysis
two quantities are equated: the signalness corresponding to the limit from the angular correlation analysis 
and the signalness of the source population of interest, given on the right hand side of Equation~\eqref{eqn:sigma}:
\begin{equation}\label{eqn:sigma}
\Sigma_{\mathrm{lim}} \overset{!}{=} \int\limits_0^{\infty}\mathrm{d}\mu\,\frac{\mathrm{d}N_{\mathrm{Sou}}}{\mathrm{d}\mu}(\mu)\frac{\mathrm{d}\Sigma}{\mathrm{d}N_{\mathrm{Sou}}}(\mu)
\end{equation}
The signalness of the source population is the integral of the 
signalness per source
$\frac{\mathrm{d}\Sigma}{\mathrm{d}N_{\mathrm{Sou}}}(\mu)$  as function of the 
source strength $\mu$ weighted with the 
source count distribution
$\frac{\mathrm{d}N_{\mathrm{Sou}}}{\mathrm{d}\mu}(\mu)$.
Solving Equation~\eqref{eqn:sigma} for parameters of an 
assumed source count distribution results in
 limits on these parameters based on the non-observation
of  angular correlations. Note that a methodically
similar analysis of gamma ray sources measured with Fermi-LAT is
presented in \cite{Cuoco:2012yf}.

The above conversion  is based on the following reasoning. We assume that the positions of
 sources in the sky are not correlated on the scale of the angular 
resolution of $\lesssim 1^\circ $ and contribute independently to the observed signalness. This results in a linear
dependence of the total signalness on the number of sources. 
As intuitively expected for an auto-correlation, the dependency of the signalness per source on the source strength is non-linear and follows a 
power law with a power index of $2$. 
Both dependencies have been verified by 
simulations as discussed  in Section~\ref{sec:intro}.

\subsection{Astrophysical flux normalization\label{subsec:HESE}}

The total normalization of the up-going diffuse astrophysical muon neutrino flux
has been measured by IceCube \cite{Aartsen:2015rwa}, and
 can be used to additionally constrain the parameters  
of a source count distribution.
For this, the total number of measured signal neutrinos expected from the source count distribution $n(\mathrm{scd})$ and the corresponding 
number expected from the observed flux $n(\mathrm{astro})$ are equated:
\begin{equation}\label{eqn:n}
n(\mathrm{scd})
=\int\limits_0^\infty\mathrm{d}\mu\,
\mu \cdot f(\gamma) \cdot
\frac{\mathrm{d}N_{\mathrm{Sou}}}{\mathrm{d}\mu}(\mu)
\overset{!}{=}n(\mathrm{astro})
=\sum\limits_{\mathrm{IC}}T^{\mathrm{IC}}\int\limits_0^\infty\mathrm{d}E\, A_{\mathrm{eff}}^{\mathrm{IC}}\cdot\frac{\mathrm{d}\Phi}{\mathrm{d}E},
\end{equation}
where $\frac{\mathrm{d}\Phi}{\mathrm{d}E}$ is the  differential 
astrophysical neutrino flux as observed by IceCube. 
The parameter values solving Equation~\eqref{eqn:n} represent 
maximum astrophysical scenarios that are consistent with the observation, 
i.e. assuming no other sources contributing to the observed flux.

\section{Application to isotropic generic sources}\label{sec:cosmo}
\subsection{Limit conversion and astrophyiscal flux normalization}
\begin{figure}[htp]
\includegraphics[width=\textwidth]{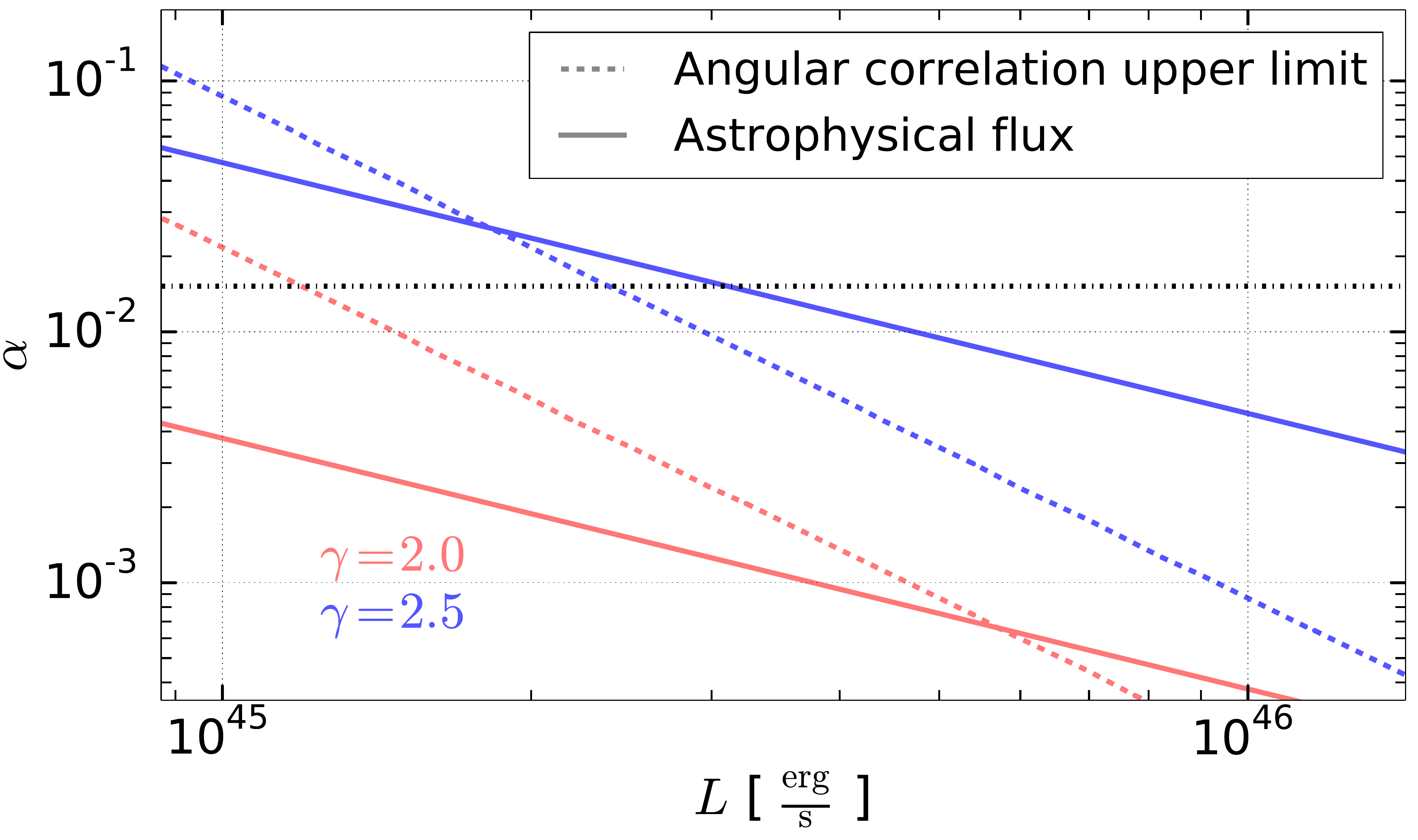}\caption{Results of the application to redshift-dependent comoving source densities;
dashed lines: IceCube angular correlation upper limits converted to $\alpha(L)$; solid lines: $\alpha(L)$ representing the observed upgoing astrophysical muon neutrino flux;
dashed-dotted line: solid angle for zenith angles between $0^\circ$ and $10^\circ$ divided by the solid angle of a hemisphere}\label{fig:cosmo}
\end{figure}

In this section, the source count distribution parametrization from Section~\ref{subsubsec:cosmo_calc} is used to solve Equations~\eqref{eqn:sigma} and~\eqref{eqn:n} for $(L,\alpha)$ for both considered energy spectra.
The solutions of Equation~\eqref{eqn:sigma} are functions $\alpha(L)$ representing the converted upper limits on these parameters.
They are shown as colored dashed lines in Figure~\ref{fig:cosmo}.

The solutions of Equation~\eqref{eqn:n} are functions $\alpha(L)$ 
representing the observed astrophysical flux for these parameters.
They are represented as colored solid lines in Figure~\ref{fig:cosmo}.

The values for $L$ at the intersections between the lines of equal colors in Figure~\ref{fig:cosmo} are the upper limits on the muon neutrino luminosity $L$ under the condition that the considered source populations produce the 
 diffuse astrophysical neutrino flux.
Therefore, all larger values for $L$ are excluded under this condition.
However, for the co-moving densities, represented by the scale factor $\alpha$, this is not the case:
The values for $\alpha$ at the intersections can be exceeded under the condition that the luminosity $L$ of each source is lower than at the intersections.
This is further discussed in Section~\ref{subsec:cosmo_conclusions}.

As introduced, a certain value of $\alpha$ can be interpreted as a fraction of all AGNs up to $z=6$, i.e. a fraction of $\frac{\mathrm{d}N_\mathrm{Sou}}{\mathrm{d}V_\mathrm{c}}\big|_\mathrm{benchmark}$.
The value for $\alpha$ at the horizontal dashed-dotted line in Figure~\ref{fig:cosmo} is a rough estimation of the blazar fraction among the AGNs represented by $\frac{\mathrm{d}N_\mathrm{Sou}}{\mathrm{d}V_\mathrm{c}}\big|_\mathrm{benchmark}$.
The estimation is based on the assumption that 
 an AGN is identified as a blazar if the angle between the AGN's jet and the viewing direction is below $10^\circ$ \cite{BL_Lac}.
For random orientations of  jet directions the corresponding 
fractional solid angle is  $1-\cos{(10^\circ)}\approx 0.015$.

\subsection{Impact of luminosity distributions}\label{subsec:dNdL}
The assumption of fixed source luminosities $L$ within a population
is not realistic.
Extended investigations could assume more realistic luminosity distributions or 
varying neutrino production efficiencies. They would, however,
involve more model parameters. 
Our values $L$ are to be considered as the `effective' $L$ of a 
population.

\begin{figure}
\includegraphics[width=\linewidth]{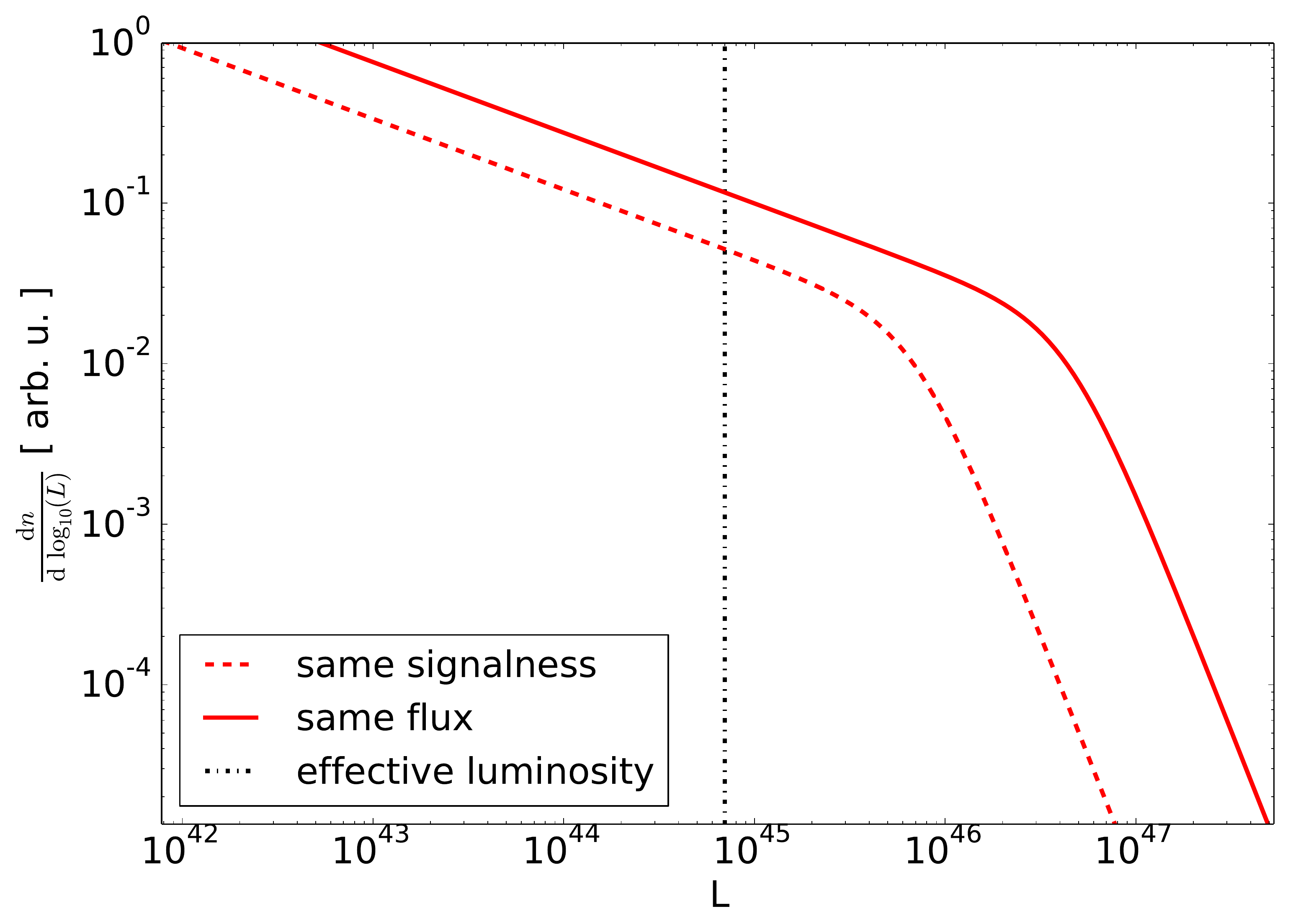}
\caption{Exemplary muon neutrino luminosity distributions with, respectively, the same mean signalness and same mean flux per source as the effective luminosity from Figure~\ref{fig:scratch_mu_z}; dashed line: distribution with the same mean signalness; solid line: distribution with the same mean flux; dashed-dotted vertical line: $L= \unit[7\cdot10^{44}]{\frac{erg}{s}}$ (same luminoity as in Figure~\ref{fig:scratch_mu_z})}
\label{fig:dNdL}
\end{figure}

In the following, a possible type of muon neutrino luminosity distributions
${\mathrm{d}n \over \mathrm{d}L }$ that correspond to a certain effective $L$,
is motivated and investigated. 
For this, a luminosity distribution based on observations of radio galaxies
at a frequency of $\unit[325]{MHz}$ is adopted from~\cite{prescott2015galaxy}.
Assuming the shape of the distribution being the same for muon neutrino luminosities,
its parametrization can be adopted for ${\mathrm{d}n \over \mathrm{d}L }$.
While the shape is determined this way, the normalization of ${\mathrm{d}n \over \mathrm{d}L }$
is arbitrary for our purposes since we only consider
the mean flux and mean signalness of the sources in the distribution to
compare it to a given effective luminosity $L$.
However, the actual normalization of the number density of sources of a population
is still solely determined by the co-moving source
density $\frac{\mathrm{d}N_\mathrm{Sou}}{\mathrm{d}V_\mathrm{c}}$. 
Thus, we obtain the two distributions shown in Figure~\ref{fig:dNdL} that differ only by a horizontal shift.
The first represents a source distribution with the same mean flux per source
as the effective luminosity $L= \unit[7\cdot10^{44}]{\frac{erg}{s}}$ used as reference.
It is shown as a solid line in Figure~\ref{fig:dNdL}.
The second represents a source distribution with the same mean signalness per source
as the effective luminosity mentioned above.
It is shown as a dashed line in Figure~\ref{fig:dNdL}.
As expected, the latter distribution has a lower mean luminosity
because the luminous sources in this distribution
are taken into account with a larger weight compared to the distribution with the same mean flux.
Depending on the quantity of interest---flux or signalness of
a certain population---for both of these distributions,
$L= \unit[7\cdot10^{44}]{\frac{erg}{s}}$ can be considered as
an effective luminosity. 
Moreover, analogously to the given example of Figure~\ref{fig:dNdL},
one can obtain distributions ${\mathrm{d}n \over \mathrm{d}L }$ using the parametrization from \cite{prescott2015galaxy}
for all effective luminosities $L$ of interest.
This allows to examine more realistic luminosity distributions corresponding to both
the observed astrophysical neutrino flux and the converted angular correlation limit,
represented by their effective luminosities $L$ as given in Figure~\ref{fig:cosmo}.

One should note, that we did not take into account the redshift dependence of
the distribution ${\mathrm{d}n \over \mathrm{d}L }$.
This is due to only dependencies at low redshifts being addressed in \cite{prescott2015galaxy},
which is insufficient for the method presented here.
Additionally, we assumed a redshift evolution of the
source densities (see e.g. Figure~\ref{fig:source_dens}), while
 the redshift dependence of the luminosity distribution in \cite{prescott2015galaxy}
holds only for the assumption of no source density evolution.
Therefore, these assumptions could not be combined easily in a self consistent way.
However, the additional correction from the explicit redshift dependence 
would only be noticable for high redshifts for which the impact on our results is low.
In conclusion, while we present our results for the simplified
case of an effective luminosity $L$, realistic luminosity distributions ${\mathrm{d}n \over \mathrm{d}L }$ 
can be mapped towards this effective luminosity as shown for the example of the parametrization given in \cite{prescott2015galaxy}.
Thus, also more sophisticated astrophysical models can be constrained
by the combination of angular correlations and the observed astrophysical neutrino flux.

\section{Application to Fermi-LAT extragalactic sources}\label{sec:Fermi}

In the following, the angular correlation limit
and the astrophysical flux normalization are interpreted in
 terms of the Fermi-LAT source 
count distribution parameters.
Then, values for a universal neutrino-to-photon ratio  
$\varepsilon_{\nu/\gamma}$ 
corresponding to the upper limit and the astrophysical flux normalization
are determined.
As a last step, 
we also determine values for $\varepsilon_{\nu/\gamma}$ corresponding to the upper limit and the astrophysical flux normalization
while varying the source count distribution parameters $\beta_1$ and $\beta_2$.
By this, we constrain the possible values that $\beta_1$ and $\beta_2$ might have for neutrinos.

\subsection{Limit conversion and astrophyiscal flux normalization}\label{subsec:Fermi_results}

\begin{figure}[htp]
\centering
\includegraphics[width=.99\textwidth]{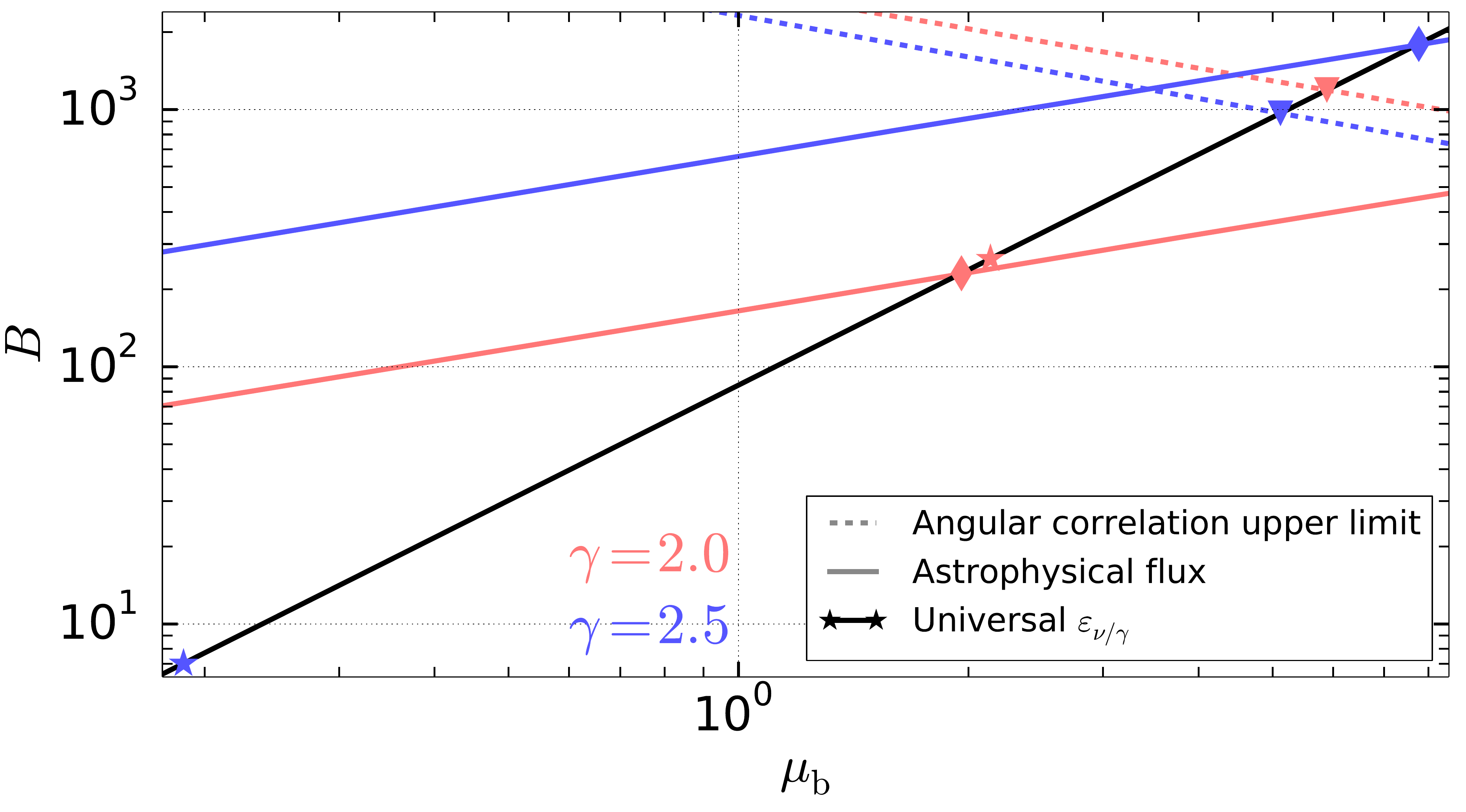}
\caption{Dashed lines: IceCube limits for both spectral indices 
converted to $B(\mu_{\mathrm{b}})$; colored solid lines: values of $(\mu_\mathrm{b},B)$ reproducing
the observed astrophysical neutrino flux; 
black line: values of $(\mu_\mathrm{b},B)$ 
that correspond to a universal value $\varepsilon_{\nu/\gamma}$;
triangles: values of $(\mu_\mathrm{b},B)$ at the IceCube limit 
that correspond to a universal value $\varepsilon_{\nu/\gamma}$; 
diamonds: values of $(\mu_\mathrm{b},B)$ that 
reproduce the observed astrophysical neutrino flux and correspond to a universal value $\varepsilon_{\nu/\gamma}$; 
asterisks: $(\mu_\mathrm{b,Fermi},B_\mathrm{Fermi})$, i.e. values such that 
$\varepsilon_{\nu/\gamma} = 1$.
}
\label{fig:results}
\end{figure}

The solution of Equation~\eqref{eqn:sigma} with $\frac{\mathrm{d}N_{\mathrm{Sou}}}{\mathrm{d}\mu}(\mu)$ from Equation~\eqref{eqn:scd} is a function $B(\mu_{\mathrm{b}})$.
It represents the upper limit from the angular correlation analysis \cite{Aartsen:2014ivk} on $B$ for each value of $\mu_{\mathrm{b}}$ and is shown as a dashed line for each spectral index in Figure~\ref{fig:results}.
The negative slopes of these lines originate from the increased 
signalness for larger $\mu_{\mathrm{b}}$ (see Equation~\eqref{eqn:sigma}) due 
to a corresponding larger non-zero integration range.
This increased signalness is compensated by lower values for $B$, 
causing the negative slope.

The astrophysical flux solutions for both neutrino energy spectra are shown as solid colored lines in Figure~\ref{fig:results}.
They differ because
 the effective area and the flux normalization in Equation~\eqref{eqn:n} are energy dependent \cite{Aartsen:2014ivk,Aartsen:2015rwa}.
Their intersections with the dashed limit lines (of the respective energy spectrum) separate the allowed region (below) from the excluded region (above) of parameter values. 
This means, a source population with larger values of $B$ or $\mu_\mathrm{b}$ cannot
produce the observed flux  of astrophysical muon 
neutrinos due to the absence of angular correlations associated with them.

\subsection{The Fermi-LAT best-fit value and the neutrino-to-photon ratio}\label{subsec:Fermi_epsilon}

The solution  $(\mu_\mathrm{b,Fermi},B_\mathrm{fermi})$, corresponding to the
special case of $\varepsilon_{\nu/\gamma} = 1$, is determined according to 
Equations~\eqref{eqn:mubfermi} and ~\eqref{eqn:Bfermi}.
It is shown as an asterisk in Figure~\ref{fig:results} for each energy spectrum.
All solutions for pairs 
of $(\mu_\mathrm{b},B)$ that correspond to
 different universal  values  $\varepsilon_{\nu/\gamma} \ne 1 $ are determined 
by Equation~\eqref{eqn:enugamma} and 
result in the black line shown in Figure~\ref{fig:results}.

The intersections of the (dashed) limit lines and the (solid colored) lines representing the observed astrophysical neutrino flux with the black line in Figure~\ref{fig:results} yield values for $\varepsilon_{\nu/\gamma}$ corresponding to the angular correlation limit and the diffuse astrophysical flux for both energy spectra in this simplified model.
These can be read off by considering that $\varepsilon_{\nu/\gamma} = \frac{\mu_\mathrm{b}}{\mu_\mathrm{b,Fermi}}$ according to Equation~\eqref{eqn:enugamma}.
By reading off the values of $\mu_\mathrm{b}$ at these intersections (triangles and diamonds in Figure~\ref{fig:results}) and $\mu_\mathrm{b,Fermi}$ (asterisks in Figure~\ref{fig:results}), one thus obtains the values for $\varepsilon_{\nu/\gamma}$.
These are given in Table~\ref{tab:results}.

\begin{table}\centering
\begin{tabular}{c c c c}
    $\gamma$ & $\varepsilon_{\nu/\gamma}$ flux normalization & $\varepsilon_{\nu/\gamma}$ correlation limit & ratio \\ \hline
    2.0 & 0.92 & 2.76 & 0.33 \\
    2.5 & 41.4 & 27.3 & 1.52 \\
\end{tabular}
\caption{Results for universal $\varepsilon_{\nu/\gamma}$; 2nd column: $\varepsilon_{\nu/\gamma}$ values assuming the observed neutrino flux; 3rd column: $\varepsilon_{\nu/\gamma}$ upper limits; 4th column: ratio between $\varepsilon_{\nu/\gamma}$ astrophysical flux value and $\varepsilon_{\nu/\gamma}$ upper limit}
\label{tab:results}
\end{table}

\subsection{Varation of the source count distribution power indices}\label{subsec:varybeta}

\begin{figure}[htp]
\begin{minipage}{.49\textwidth}
\centering $\gamma=2.0$
\end{minipage}
\begin{minipage}{.49\textwidth}
\centering $\gamma=2.5$
\end{minipage}
\centering
\begin{subfigure}{.5\textwidth}
  \centering
\includegraphics[width=\textwidth]{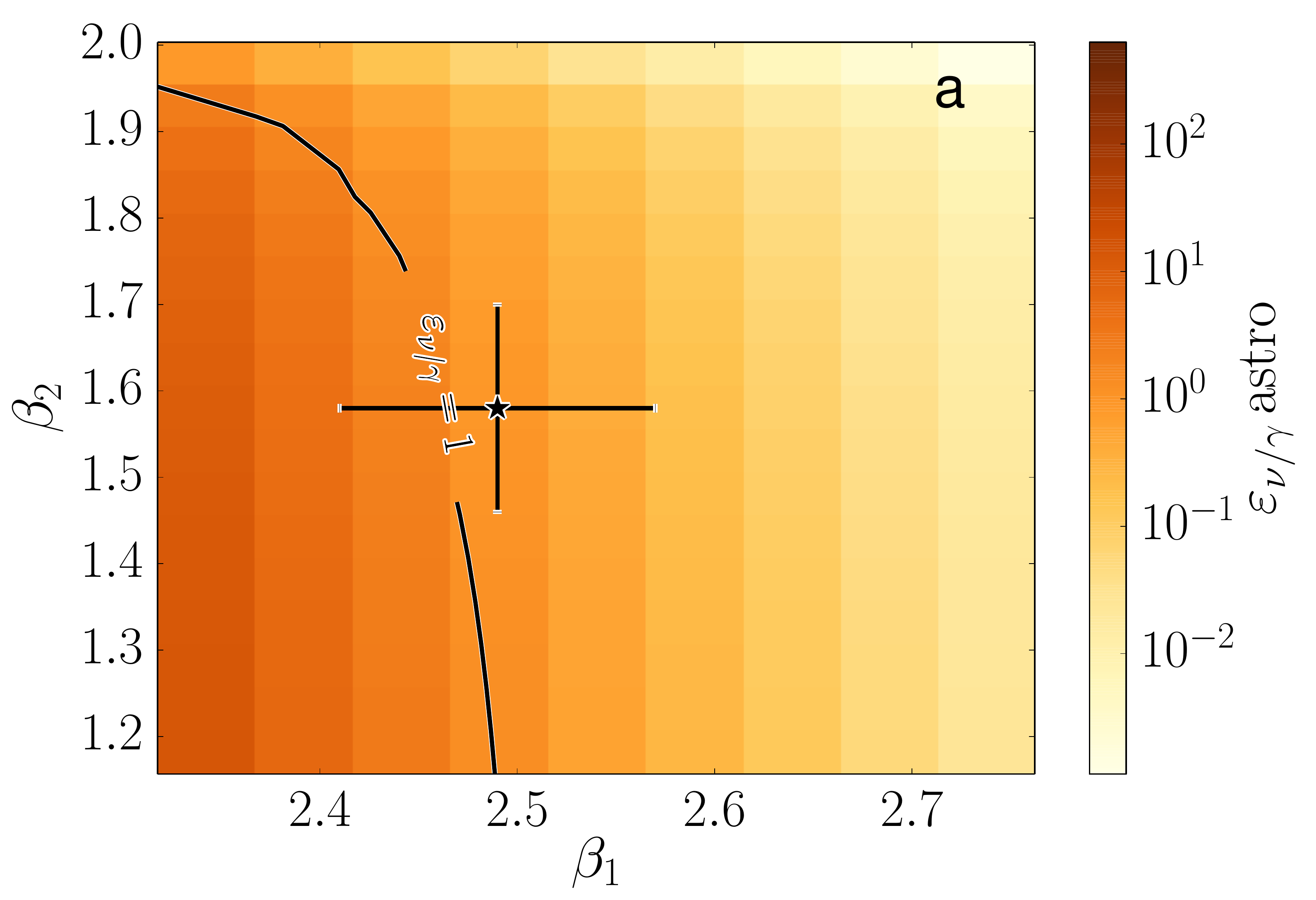}
  \refstepcounter{subfigure}\label{fig:gamma_2_pred}
\end{subfigure}%
\begin{subfigure}{.5\textwidth}
  \centering
\includegraphics[width=\textwidth]{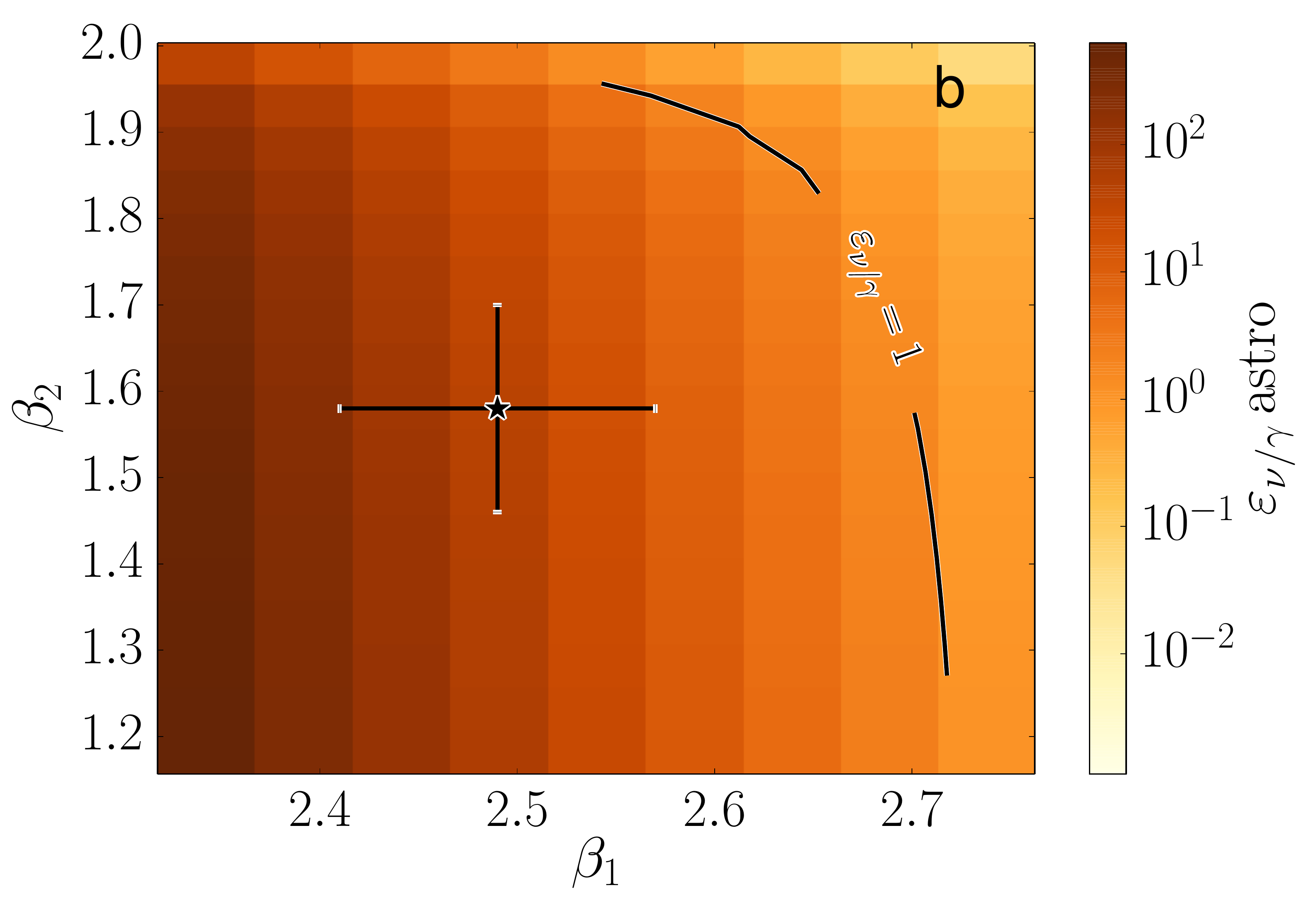}
    \refstepcounter{subfigure}\label{fig:gamma_25_pred}
\end{subfigure}\vspace{-11pt}
\begin{subfigure}{.5\textwidth}
  \centering
\includegraphics[width=\textwidth]{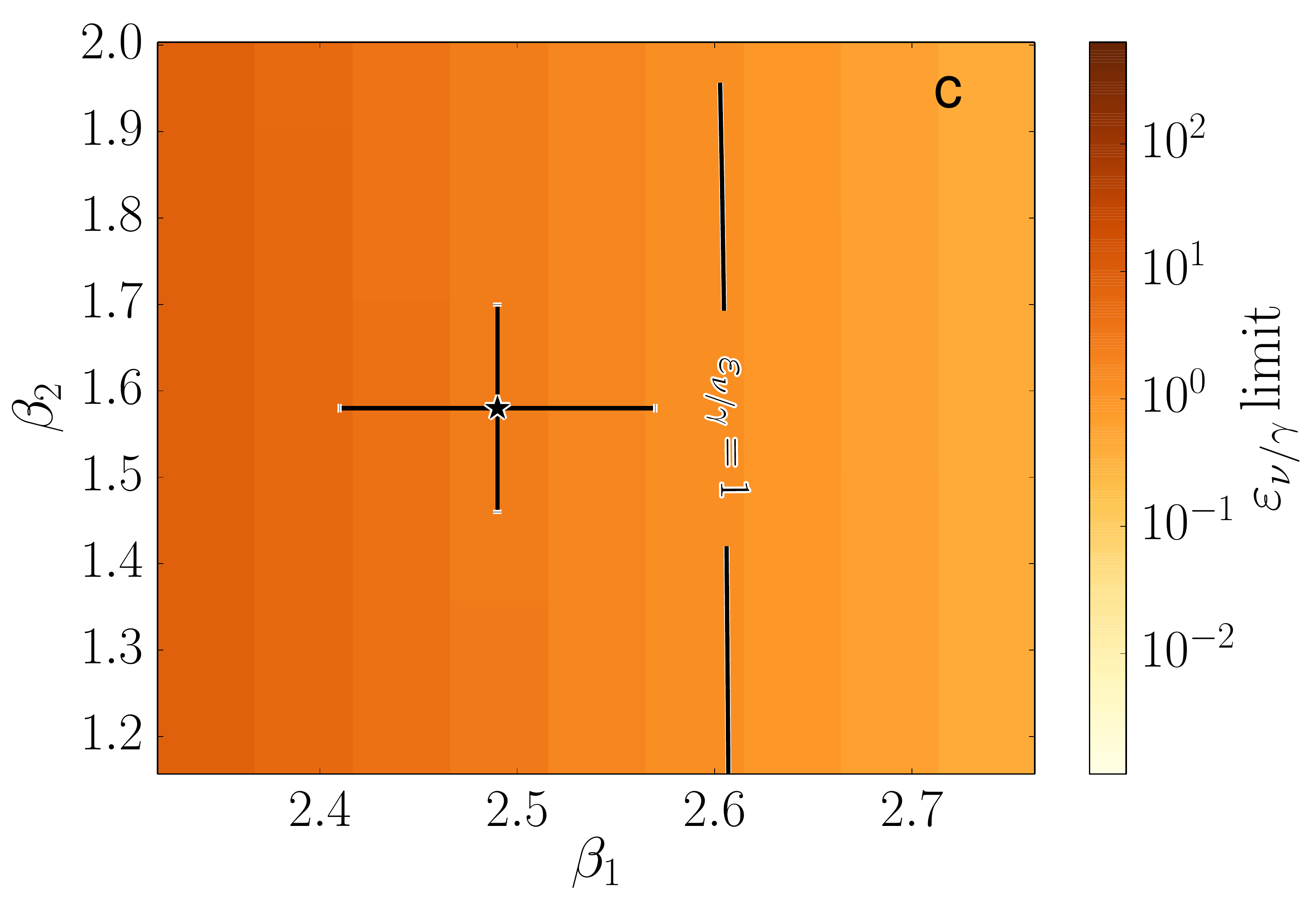}
    \refstepcounter{subfigure}\label{fig:gamma_2_lim}
\end{subfigure}%
\begin{subfigure}{.5\textwidth}
  \centering
\includegraphics[width=\textwidth]{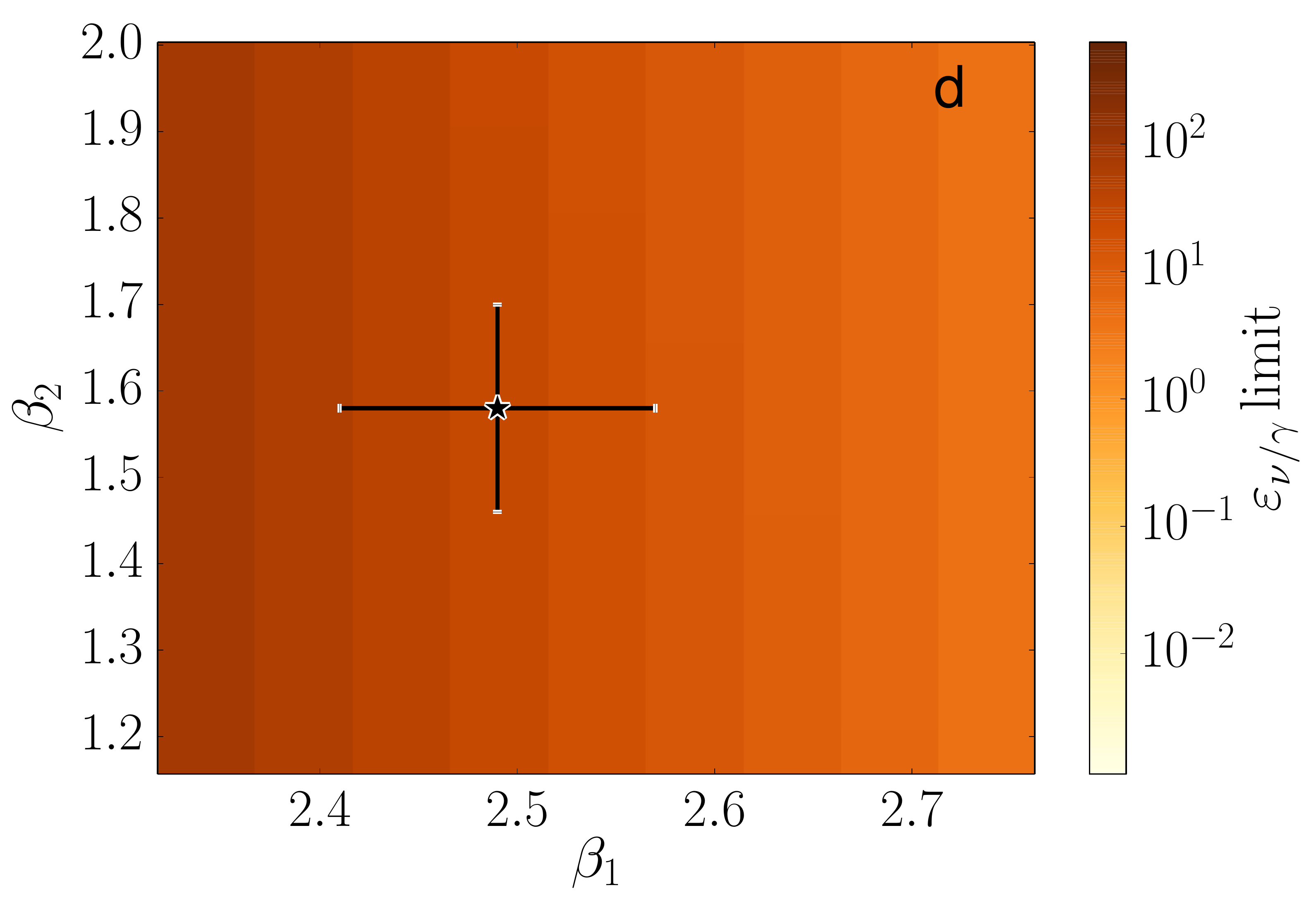}
    \refstepcounter{subfigure}\label{fig:gamma_25_lim}
\end{subfigure}\vspace{-11pt}
\begin{subfigure}{.5\textwidth}
  \centering
\includegraphics[width=\textwidth]{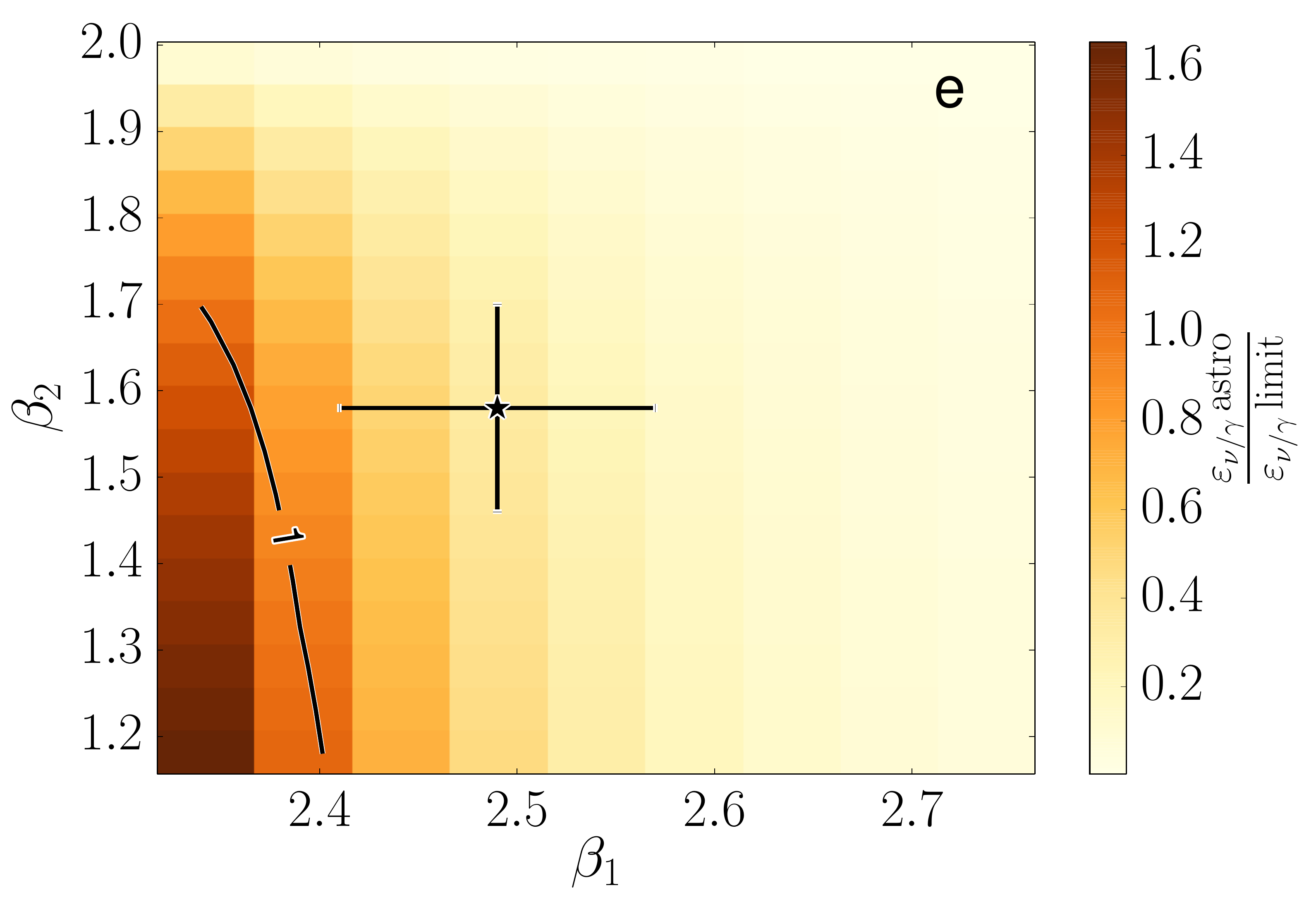}
    \refstepcounter{subfigure}\label{fig:gamma_2_ratio}
\end{subfigure}%
\begin{subfigure}{.5\textwidth}
  \centering
\includegraphics[width=\textwidth]{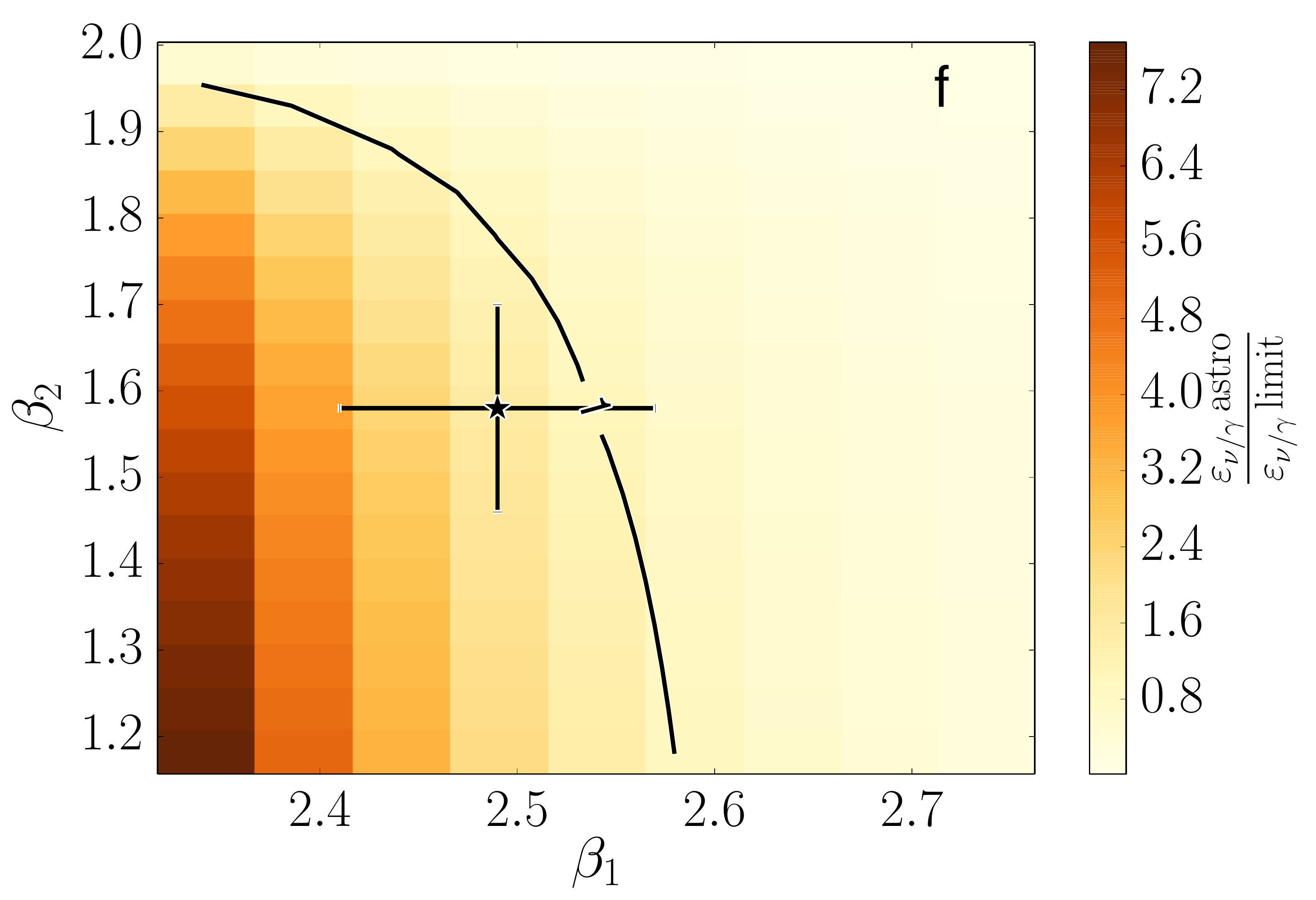}
  \refstepcounter{subfigure}\label{fig:gamma_25_ratio}
\end{subfigure}\vspace{-11pt}
\caption{a, b: universal neutrino-to-photon ratios $\varepsilon_{\nu/\gamma}$ for source populations that correspond to the observed astrophysical neutrino flux for different power-indices $\beta_1$, $\beta_2$ (s. Equation \protect\eqref{eqn:scd}); c, d: converted correlation limits on an 
universal $\varepsilon_{\nu/\gamma}$ for different powers $\beta_1$, $\beta_2$; 
e, f: ratios between  $\varepsilon_{\nu/\gamma}$ flux prediction and the $\varepsilon_{\nu/\gamma}$ limit; asterisk with error bars: $\beta_1$, $\beta_2$ and uncertainties from Fermi-LAT \cite{Fermi}; black line in a-c: $\varepsilon_{\nu/\gamma}= 1$; black line in e, f: ratio $= 1$}
\label{fig:scan}
\end{figure}

The used source count distribution parametrization (Equation~\eqref{eqn:scd}) can be generalized by varying the powers
 $\beta_1$ and $\beta_2$ and repeating the procedure from Sections~\ref{subsec:Fermi_results} and \ref{subsec:Fermi_epsilon}.
In Figure~\ref{fig:gamma_2_pred} and~\ref{fig:gamma_25_pred},the results are shown in terms of the $\varepsilon_{\nu/\gamma}$ astrophysical flux values, while in Figure~\ref{fig:gamma_2_lim} and~\ref{fig:gamma_25_lim} the $\varepsilon_{\nu/\gamma}$ upper limit is shown for the neutrino energy spectra with $\gamma=2.0$ and $\gamma=2.5$.
Finally, the ratios between the $\varepsilon_{\nu/\gamma}$ astrophysical flux value and the $\varepsilon_{\nu/\gamma}$ upper limit are shown in Figure~\ref{fig:gamma_2_ratio} and~\ref{fig:gamma_25_ratio}.
Ratios larger than $1$ indicate that the astrophysical neutrino flux is excluded to originate purely from the corresponding 
source population  with 90 \% C.L. based on the non-observation of angular correlations and assuming a universal $\varepsilon_{\nu/\gamma}$ in the considered energy ranges.

\section{Conclusions}\label{sec:conclusions}

\subsection{Angular correlations from  generic AGN-type sources}\label{subsec:cosmo_conclusions}

The discussion of results from Section~\ref{sec:cosmo} is based on Figure
\ref{fig:cosmo}.
The angular correlation analysis constrains the allowed parameter space to the region below the dashed lines.
However, also the observed diffuse astrophysical neutrino flux reflects an upper limit on the maximum allowed 
contribution by these sources and also parameter regions above the solid lines are excluded.

For a spectral index of $\gamma =2.0$, the constraints by the non-observation of angular correlations are weaker than the observed flux except for very large muon neutrino luminosities of the sources above $L \simeq \unit[6 \cdot 10^{45}]{erg \over s} $ in the considered energy range.
For such sources, an angular correlation should have been observed and  the fraction $\alpha $ of these sources to the total population of AGN is constrained.
The fraction $\alpha $ would be at least
a factor 20 smaller than the estimated fraction of blazars.
For source luminosities well below $ \unit[10^{45}]{erg \over s} $,
 where the population fraction of blazars coincides with the observed flux, the angular correlation analysis does not provide additional constraints.

The situation is different for $\gamma=2.5 $. Here, the non-observation of 
angular correlations excludes luminosities above $L \simeq \unit[2\cdot 10^{45}]{erg \over s} $ stronger than the constraint by the flux normalization does.
The allowed parameter region would include an AGN fraction corresponding to the estimation for blazars, if their muon neutrino luminosity would reach such large values. 
Obviously, an improved exposure
could allow to positively detect such sources.
On the other hand the angular correlation analysis excludes that blazars 
are fully responsible for the observed flux as the required source 
luminosities $L \simeq \unit [3  \cdot 10^{45}]{erg \over s} $ are excluded.

These conclusions depend on idealized assumptions as for
example the assumption of one effective luminosity $L$ for a whole source population.
In Section~\ref{subsec:dNdL} we showed that the given
values for $L$ can be interpreted in terms of specific luminosity distributions ${\mathrm{d}n \over \mathrm{d}L }$
and showed examples of these distributions
that correspond to a certain value of $L$.
While we focused on radio sources to motivate this exemplary ${\mathrm{d}n \over \mathrm{d}L }$,
one can easily examine other parametrizations or
types of luminosity distributions and interpret IceCube's muon neutrino
angular correlation limit and observed astrophysical flux
in terms of these parametrizations using the method we present in this work.

In order to further interpret the effective luminosities $L$ one can also compare
them to AGN disk luminosities $L_\mathrm{disk}$ estimated in \cite{falcke1994jet}.
From this work, we use the $L_\mathrm{disk}$ of AGNs classified as radio loud and 
assume the corresponding jet luminosities $L_\mathrm{jet}$ to be 10\% of $L_\mathrm{disk}$.
As our value $L$ is the effective luminosity of a population, we take 
$<L_\mathrm{jet}>\approx\unit[5\cdot10^{45}]{\frac{erg}{s}}$ also as an estimate
for the effective luminosity of jets.
However, one should note that this estimate is conservative since the
angular correlation analysis gains in sensitivity per source $\propto L^2$.
Weighting sources according to $L^2$ leads to an effective jet
luminosity $\sqrt{<L_\mathrm{jet}^2>}\approx\unit[9\cdot10^{45}]{\frac{erg}{s}}$ which is well above
the given limit on the effective luminosity.
This means that it is possible to interpret our result in terms of
an upper limit on $f_{\mathrm{jet,}\nu}$, which is the fraction of luminosity transferred from radio loud quasar jets into neutrinos.
Specifically, for the aforementioned case that blazars constitute the
detected neutrino flux with an energy spectrum of $\gamma=2.5$, the upper limit on the
effective luminosity was found to be $L \simeq \unit [3  \cdot 10^{45}]{erg \over s} $.
Thus, applying the effective jet luminosity of $\unit [9  \cdot 10^{45}]{erg \over s}$ we obtain 
a constraint of $f_{\mathrm{jet,}\nu}<33\%$ for the fraction of the AGN luminosity emitted in the neutrino channel. 
Although, this is only a rough estimate, one should note that using this method future angular correlation analysis in IceCube
might contribute significantly to constraining this fraction for more sophisticated scenarios.

\subsection{Angular correlations from Fermi-LAT extragalactic sources}

Unlike the modelling of generic AGN-type sources, the result here is based
on an empirically observed source count distribution motivated by the observation of extragalactic high-energy gamma ray sources by Fermi-LAT. Therefore, the interpretation
requires the assumption of a neutrino-to-photon ratio $\varepsilon_{\nu/\gamma}$. In order to simplify the interpretation
 we assume this ratio to be universal for all sources.

Starting with the results in Figure\ \ref{fig:results} and Table\ \ref{tab:results} we see that
for hard spectrum sources with $\gamma=2.0$, the observed astrophysical neutrino flux corresponds to a lower $\varepsilon_{\nu/\gamma}$ value than the $\varepsilon_{\nu/\gamma}$ limit from the correlation analysis.
Hence, for this energy spectrum, the sources from the Fermi-LAT high-latitude survey are not excluded to be the origin of the astrophysical neutrino flux under the stated assumptions.
Furthermore, the flux normalization results in a required
neutrino-to-photon ratio close to the generic value  of $\varepsilon_{\nu/\gamma} \simeq 1$.  An improved sensitivity of about a factor $3$ for the correlation analysis is required to to test this value. This seems
well feasible with future data of IceCube.

For a softer spectrum with $\gamma=2.5$, the opposite is the case: The required value $\varepsilon_{\nu/\gamma}$ for the astrophysical flux normalization 
is excluded by the $\varepsilon_{\nu/\gamma}$ limit from not observing angular correlations.
This means that the astrophysical flux is excluded to be produced exclusively by sources distributed according to the Fermi-LAT motivated source count distribution parametrization for this spectrum.
Furthermore, those sources would need to have neutrino-to-photon ratios of 
$\simeq 40$.
Thus, besides their apparent absence, such high ratios would also need to be explained theoretically.

We note that the assumption of a universal  
 $\varepsilon_{\nu/\gamma} \approx 1 $ is not a robust assumption and is considered as a benchmark, only.
The initial value  $\varepsilon_{\nu/\gamma} $ strongly 
depends on the specific hadronic production mechanism, the density
of the medium,  as well as
 energy losses or acceleration of intermediate  particles \cite{Gaisser:1994yf,Becker:2007sv,Klein:2012ug}.
Then, depending on the optical depth of the sources, absorption  of photons  would lead to larger ratios \cite{Gaisser:1994yf}.
However, during propagation the muon neutrino flux is also modified 
due to oscillations to other flavors  (see e.g.\ \cite{Anchordoqui:2013dnh}).
In case of Fermi-Lat,  the determined ratio depends on
observations at largely different energy scales. It is questionable
whether all sources that contribute to the Fermi-LAT source count distribution are actually dominated by photons from hadronic interactions.
A strong leptonic contribution could result in substantially smaller  $\varepsilon_{\nu/\gamma}$ values.
As another effect, the absorption of photons during propagation is weak 
for the Fermi-LAT energy range  affecting only the most distant sources.
For the limits from the angular correlation analysis, this effect can be neglected as these are dominantly
affected by the closest bright sources.
Still, it would modify the total flux normalization and hence, for a fixed flux normalization, the exclusion power with respect to Fermi-LAT would be reduced.
However, the obtained results provide constraints of the properties of astrophysical neutrino sources under these simplified assumptions. 
By including the effects discussed above, one can modify the results in order to obtain more specific constraints in terms of astrophysical 
source properties. For such specific modelling, the methods introduced in this work are still applicable in the same way.

Motivated by these systematic uncertainties,  the studies
of variations  of $\beta_1$ and $\beta_2$ (s. Equations~\eqref{eqn:scd_original} and~\eqref{eqn:scd}) reveal several insights:
First, $\beta_1$ plays a strong role for both, the $\varepsilon_{\nu/\gamma}$ astrophysical flux value and the $\varepsilon_{\nu/\gamma}$ limit because the source count distribution depends strongly on $\beta_1$ for all source strengths $\mu$.
Second, the $\varepsilon_{\nu/\gamma}$ limit is almost independent of $\beta_2$ while the $\varepsilon_{\nu/\gamma}$ astrophysical flux value noticeably depends on $\beta_2$. This is obvious as
$\beta_2$ only affects the source count distribution $\frac{\mathrm{d}N_{\mathrm{Sou}}}{\mathrm{d}\mu}$ for source strengths below the break $\mu_{\mathrm{b}}$ (s. Equation~\eqref{eqn:scd}), i.e. `weak' sources.

The quantities used for the $\varepsilon_{\nu/\gamma}$ limit and the $\varepsilon_{\nu/\gamma}$ astrophysical flux value are the signalness $\Sigma$ and the number of neutrinos $n(\mathrm{scd})$ from the tested source count distribution.
This leads to the conclusion that the sources brighter than $\mu_{\mathrm{b}}$, i.e. `strong' sources, are the signalness dominating sources while weak sources affect only $n(\mathrm{scd})$ and not the signalness $\Sigma$.
This is a direct consequence of the definitions of $n(\mathrm{scd})$ (Equation~\eqref{eqn:n}) and $\Sigma$ (Equation~\eqref{eqn:sigma}) which depend on different powers of the source strength $\mu$.

A ratio between a value for $\varepsilon_{\nu/\gamma}$ and the $\varepsilon_{\nu/\gamma}$ limit larger than 1 is excluded with 90 \% C.L.
Thus, for both energy spectra, the areas to the bottom left from the black lines in Figures~\ref{fig:gamma_2_ratio} and~\ref{fig:gamma_25_ratio} are excluded.
For $\gamma=2.5$, where the hypothesis with the benchmark values for $\beta_1$ and $\beta_2$ is excluded, its uncertainty interval reaches into the allowed region.


\section{Summary}

We have developed a method to interpret the results from analyses of 
angular correlations in IceCube muon neutrino data 
 in terms of astrophysical scenarios. In addition, the
 observed astrophysical neutrino flux can be introduced as a boundary condition.
We have shown that already with early data from the partly installed 
IceCube detector  astrophysical scenarios can be constrained.
This is especially the case for soft energy spectra.
We expect a substantially improved sensitivity  once results 
for the angular correlation with the full IceCube detector become available.

\section*{Acknowledgements}

This work is supported by the Federal Ministry of Education and Research (BMBF), the Helmholtz Alliance for Astroparticle Physics (HAP) and the German Research Foundation (DFG).
We thank Markus Ahlers, Jan Auffenberg, Alessandro Cuoco,  Julien Lesgourgues, Leif R\"adel and Julia Tjus for the valuable discussions.

\section*{References}







\bibliographystyle{model5-names}\biboptions{numbers}


\bibliography{mybibfile,references}

\begin{thebibliography}{28}
\expandafter\ifx\csname natexlab\endcsname\relax\def\natexlab#1{#1}\fi
\providecommand{\url}[1]{\texttt{#1}}
\providecommand{\href}[2]{#2}
\providecommand{\path}[1]{#1}
\providecommand{\DOIprefix}{doi:}
\providecommand{\ArXivprefix}{arXiv:}
\providecommand{\URLprefix}{URL: }
\providecommand{\Pubmedprefix}{pmid:}
\providecommand{\doi}[1]{\href{http://dx.doi.org/#1}{\path{#1}}}
\providecommand{\Pubmed}[1]{\href{pmid:#1}{\path{#1}}}
\providecommand{\bibinfo}[2]{#2}
\ifx\xfnm\relax \def\xfnm[#1]{\unskip,\space#1}\fi
\bibitem[{Aartsen et~al.(2013)}]{Aartsen:2013jdh}
\bibinfo{author}{Aartsen, M.~G.} et~al. (\bibinfo{collaboration}{IceCube})
  (\bibinfo{year}{2013}).
\newblock \bibinfo{title}{{Evidence for High-Energy Extraterrestrial Neutrinos
  at the IceCube Detector}}.
\newblock {\it \bibinfo{journal}{Science}\/},  {\it \bibinfo{volume}{342}\/},
  \bibinfo{pages}{1242856}. \DOIprefix\doi{10.1126/science.1242856}.
  \href{http://arxiv.org/abs/1311.5238}{\tt arXiv:1311.5238}.
\bibitem[{Aartsen et~al.(2014{\natexlab{a}})}]{Aartsen:2014gkd}
\bibinfo{author}{Aartsen, M.~G.} et~al. (\bibinfo{collaboration}{IceCube})
  (\bibinfo{year}{2014}{\natexlab{a}}).
\newblock \bibinfo{title}{{Observation of High-Energy Astrophysical Neutrinos
  in Three Years of IceCube Data}}.
\newblock {\it \bibinfo{journal}{Phys. Rev. Lett.}\/},  {\it
  \bibinfo{volume}{113}\/}, \bibinfo{pages}{101101}.
  \DOIprefix\doi{10.1103/PhysRevLett.113.101101}.
  \href{http://arxiv.org/abs/1405.5303}{\tt arXiv:1405.5303}.
\bibitem[{Aartsen et~al.(2014{\natexlab{b}})}]{Aartsen:2014cva}
\bibinfo{author}{Aartsen, M.~G.} et~al. (\bibinfo{collaboration}{IceCube})
  (\bibinfo{year}{2014}{\natexlab{b}}).
\newblock \bibinfo{title}{{Searches for Extended and Point-like Neutrino
  Sources with Four Years of IceCube Data}}.
\newblock {\it \bibinfo{journal}{Astrophys. J.}\/},  {\it
  \bibinfo{volume}{796}\/}, \bibinfo{pages}{109}.
  \DOIprefix\doi{10.1088/0004-637X/796/2/109}.
  \href{http://arxiv.org/abs/1406.6757}{\tt arXiv:1406.6757}.
\bibitem[{Aartsen et~al.(2015{\natexlab{a}})}]{Aartsen:2015rwa}
\bibinfo{author}{Aartsen, M.~G.} et~al. (\bibinfo{collaboration}{IceCube})
  (\bibinfo{year}{2015}{\natexlab{a}}).
\newblock \bibinfo{title}{{Evidence for Astrophysical Muon Neutrinos from the
  Northern Sky with IceCube}}.
\newblock {\it \bibinfo{journal}{Phys. Rev. Lett.}\/},  {\it
  \bibinfo{volume}{115}\/}, \bibinfo{pages}{081102}.
  \DOIprefix\doi{10.1103/PhysRevLett.115.081102}.
  \href{http://arxiv.org/abs/1507.04005}{\tt arXiv:1507.04005}.
\bibitem[{Aartsen et~al.(2015{\natexlab{b}})}]{Aartsen:2014ivk}
\bibinfo{author}{Aartsen, M.~G.} et~al. (\bibinfo{collaboration}{IceCube})
  (\bibinfo{year}{2015}{\natexlab{b}}).
\newblock \bibinfo{title}{{Searches for small-scale anisotropies from neutrino
  point sources with three years of IceCube data}}.
\newblock {\it \bibinfo{journal}{Astropart. Phys.}\/},  {\it
  \bibinfo{volume}{66}\/}, \bibinfo{pages}{39--52}.
  \DOIprefix\doi{10.1016/j.astropartphys.2015.01.001}.
  \href{http://arxiv.org/abs/1408.0634}{\tt arXiv:1408.0634}.
\bibitem[{Abdo et~al.(2010)}]{Fermi}
\bibinfo{author}{Abdo, A.~A.} et~al. (\bibinfo{year}{2010}).
\newblock \bibinfo{title}{{The {Fermi-LAT} High-Latitude Survey: Source Count
  Distributions and the Origin of the Extragalactic Diffuse Background}}.
\newblock {\it \bibinfo{journal}{Astrophys. J.}\/},  {\it
  \bibinfo{volume}{720}\/}, \bibinfo{pages}{435}.
  \DOIprefix\doi{10.1088/0004-637X/720/1/435}.
  \href{http://arxiv.org/abs/1003.0895}{\tt arXiv:1003.0895}.
\bibitem[{Achterberg et~al.(2006)}]{IC}
\bibinfo{author}{Achterberg, A.} et~al. (\bibinfo{collaboration}{IceCube})
  (\bibinfo{year}{2006}).
\newblock \bibinfo{title}{First year performance of the {IceCube} neutrino
  telescope}.
\newblock {\it \bibinfo{journal}{Astropart. Phys.}\/},  {\it
  \bibinfo{volume}{26}\/}, \bibinfo{pages}{155--173}.
  \DOIprefix\doi{10.1016/j.astropartphys.2006.06.007}.
  \href{http://arxiv.org/abs/astro-ph/0604450}{\tt arXiv:astro-ph/0604450}.
\bibitem[{Ajello et~al.(2014)}]{BL_Lac}
\bibinfo{author}{Ajello, M.} et~al. (\bibinfo{year}{2014}).
\newblock \bibinfo{title}{{The Cosmic Evolution of {Fermi} {BL} {Lacertae}
  Objects}}.
\newblock {\it \bibinfo{journal}{Astrophys. J.}\/},  {\it
  \bibinfo{volume}{780}\/}, \bibinfo{pages}{73}.
  \DOIprefix\doi{10.1088/0004-637X/780/1/73}.
  \href{http://arxiv.org/abs/1310.1006}{\tt arXiv:1310.1006}.
\bibitem[{Anchordoqui et~al.(2014)}]{Anchordoqui:2013dnh}
\bibinfo{author}{Anchordoqui, L.~A.} et~al. (\bibinfo{year}{2014}).
\newblock \bibinfo{title}{{Cosmic Neutrino Pevatrons: A Brand New Pathway to
  Astronomy, Astrophysics, and Particle Physics}}.
\newblock {\it \bibinfo{journal}{JHEAp}\/},  {\it \bibinfo{volume}{1--2}\/},
  \bibinfo{pages}{1--30}. \DOIprefix\doi{10.1016/j.jheap.2014.01.001}.
  \href{http://arxiv.org/abs/1312.6587}{\tt arXiv:1312.6587}.
\bibitem[{Asada et~al.(2008)Asada, Doi, Nakamura, Nagai and Inoue}]{M87}
\bibinfo{author}{Asada, K.}, \bibinfo{author}{Doi, A.},
  \bibinfo{author}{Nakamura, M.}, \bibinfo{author}{Nagai, H.}, and
  \bibinfo{author}{Inoue, M.} (\bibinfo{year}{2008}).
\newblock \bibinfo{title}{{EVN} observations of {M} 87}.
\newblock \bibinfo{publisher}{9th EVN Symposium, Bologna}.
\bibitem[{Becker(2008)}]{Becker:2007sv}
\bibinfo{author}{Becker, J.~K.} (\bibinfo{year}{2008}).
\newblock \bibinfo{title}{{High-Energy Neutrinos in the Context of
  Multimessenger Physics}}.
\newblock {\it \bibinfo{journal}{Phys. Rept.}\/},  {\it
  \bibinfo{volume}{458}\/}, \bibinfo{pages}{173--246}.
  \DOIprefix\doi{10.1016/j.physrep.2007.10.006}.
  \href{http://arxiv.org/abs/0710.1557}{\tt arXiv:0710.1557}.
\bibitem[{Becker et~al.(2007)}]{becker2007astrophysical}
\bibinfo{author}{Becker, J.~K.} et~al. (\bibinfo{year}{2007}).
\newblock \bibinfo{title}{{Astrophysical Implications of High Energy Neutrino
  Limits}}.
\newblock {\it \bibinfo{journal}{Astropart. Phys.}\/},  {\it
  \bibinfo{volume}{28}\/}, \bibinfo{pages}{98--118}.
  \DOIprefix\doi{10.1016/j.astropartphys.2007.04.007}.
  \href{http://arxiv.org/abs/astro-ph/0607427}{\tt arXiv:astro-ph/0607427}.
\bibitem[{Brusa et~al.(2009)}]{brusa}
\bibinfo{author}{Brusa, M.} et~al. (\bibinfo{year}{2009}).
\newblock \bibinfo{title}{{High-Redshift Quasars in the COSMOS Survey: The
  Space Density of z $>$ 3 X-ray Selected QSOs}}.
\newblock {\it \bibinfo{journal}{Astrophys. J.}\/},  {\it
  \bibinfo{volume}{693}\/}, \bibinfo{pages}{8}.
  \DOIprefix\doi{10.1088/0004-637X/693/1/8}.
  \href{http://arxiv.org/abs/0809.2513}{\tt arXiv:0809.2513}.
\bibitem[{Cuoco et~al.(2012)Cuoco, Komatsu and Siegal-Gaskins}]{Cuoco:2012yf}
\bibinfo{author}{Cuoco, A.}, \bibinfo{author}{Komatsu, E.}, and
  \bibinfo{author}{Siegal-Gaskins, J.~M.} (\bibinfo{year}{2012}).
\newblock \bibinfo{title}{{Joint Anisotropy and Source Count Constraints on the
  Contribution of Blazars to the Diffuse Gamma-Ray Background}}.
\newblock {\it \bibinfo{journal}{Phys. Rev.}\/},  {\it
  \bibinfo{volume}{D86}\/}, \bibinfo{pages}{063004}.
  \DOIprefix\doi{10.1103/PhysRevD.86.063004}.
  \href{http://arxiv.org/abs/1202.5309}{\tt arXiv:1202.5309}.
\bibitem[{Eichmann et~al.(2012)Eichmann, Schlickeiser and
  Rhode}]{eichmann2012differences}
\bibinfo{author}{Eichmann, B.}, \bibinfo{author}{Schlickeiser, R.}, and
  \bibinfo{author}{Rhode, W.} (\bibinfo{year}{2012}).
\newblock \bibinfo{title}{{Differences of Leptonic and Hadronic Radiation
  Production in Flaring Blazars}}.
\newblock {\it \bibinfo{journal}{Astrophys. J.}\/},  {\it
  \bibinfo{volume}{749}\/}, \bibinfo{pages}{155}.
  \DOIprefix\doi{10.1088/0004-637X/749/2/155}.
\bibitem[{Falcke et~al.(1995)Falcke, Malkan and Biermann}]{falcke1994jet}
\bibinfo{author}{Falcke, H.}, \bibinfo{author}{Malkan, M.~A.}, and
  \bibinfo{author}{Biermann, P.~L.} (\bibinfo{year}{1995}).
\newblock \bibinfo{title}{{The Jet-Disk Symbiosis. II. Interpreting the
  Radio/UV Correlations in Quasars}}.
\newblock {\it \bibinfo{journal}{A\&A}\/},  {\it \bibinfo{volume}{298}\/},
  \bibinfo{pages}{375}. \href{http://arxiv.org/abs/astro-ph/9411100}{\tt
  arXiv:astro-ph/9411100}.
\bibitem[{Gaisser et~al.(1995)Gaisser, Halzen and Stanev}]{Gaisser:1994yf}
\bibinfo{author}{Gaisser, T.~K.}, \bibinfo{author}{Halzen, F.}, and
  \bibinfo{author}{Stanev, T.} (\bibinfo{year}{1995}).
\newblock \bibinfo{title}{{Particle Astrophysics with High-Energy Neutrinos}}.
\newblock {\it \bibinfo{journal}{Phys. Rept.}\/},  {\it
  \bibinfo{volume}{258}\/}, \bibinfo{pages}{173--236}.
  \DOIprefix\doi{10.1016/0370-1573(95)00003-Y}.
  \href{http://arxiv.org/abs/hep-ph/9410384}{\tt arXiv:hep-ph/9410384}.
\bibitem[{Hiroi et~al.(2012)Hiroi, Ueda, Akiyama and Watson}]{hiroi}
\bibinfo{author}{Hiroi, K.}, \bibinfo{author}{Ueda, Y.},
  \bibinfo{author}{Akiyama, M.}, and \bibinfo{author}{Watson, M.~G.}
  (\bibinfo{year}{2012}).
\newblock \bibinfo{title}{{Comoving Space Density and Obscured Fraction of
  High-Redshift Active Galactic Nuclei in the {Subaru/XMM-Newton} Deep
  Survey}}.
\newblock {\it \bibinfo{journal}{Astrophys. J.}\/},  {\it
  \bibinfo{volume}{758}\/}, \bibinfo{pages}{49}.
  \DOIprefix\doi{10.1086/429361}. \href{http://arxiv.org/abs/1208.5050}{\tt
  arXiv:1208.5050}.
\bibitem[{Klein et~al.(2013)Klein, Mikkelsen and Becker~Tjus}]{Klein:2012ug}
\bibinfo{author}{Klein, S.~R.}, \bibinfo{author}{Mikkelsen, R.~E.}, and
  \bibinfo{author}{Becker~Tjus, J.} (\bibinfo{year}{2013}).
\newblock \bibinfo{title}{{Muon Acceleration in Cosmic-Ray Sources}}.
\newblock {\it \bibinfo{journal}{Astrophys. J.}\/},  {\it
  \bibinfo{volume}{779}\/}, \bibinfo{pages}{106}.
  \DOIprefix\doi{10.1088/0004-637X/779/2/106}.
  \href{http://arxiv.org/abs/1208.2056}{\tt arXiv:1208.2056}.
\bibitem[{M{\"u}cke and Protheroe(2001)}]{mucke2001proton}
\bibinfo{author}{M{\"u}cke, A.}, and \bibinfo{author}{Protheroe, R.}
  (\bibinfo{year}{2001}).
\newblock \bibinfo{title}{{A Proton Synchrotron Blazar Model for Flaring in
  Markarian 501}}.
\newblock {\it \bibinfo{journal}{Astropart. Phys.}\/},  {\it
  \bibinfo{volume}{15}\/}, \bibinfo{pages}{121--136}.
  \DOIprefix\doi{10.1016/S0927-6505(00)00141-9}.
  \href{http://arxiv.org/abs/astro-ph/0004052}{\tt arXiv:astro-ph/0004052}.
\bibitem[{Murase et~al.(2015)Murase, Guetta and Ahlers}]{murase2015hidden}
\bibinfo{author}{Murase, K.}, \bibinfo{author}{Guetta, D.}, and
  \bibinfo{author}{Ahlers, M.} (\bibinfo{year}{2015}).
\newblock \bibinfo{title}{{Hidden Cosmic-Ray Accelerators as an Origin of
  TeV-PeV Cosmic Neutrinos}}.
\newblock {\it \bibinfo{journal}{Phys. Rev. Lett.}\/},  {\it
  \bibinfo{volume}{116}\/}, \bibinfo{pages}{071101}.
  \DOIprefix\doi{10.1103/PhysRevLett.116.071101}.
  \href{http://arxiv.org/abs/1509.00805}{\tt arXiv:1509.00805}.
\bibitem[{Olive et~al.(2014)}]{PDG}
\bibinfo{author}{Olive, K.~A.} et~al. (\bibinfo{collaboration}{Particle Data
  Group}) (\bibinfo{year}{2014}).
\newblock \bibinfo{title}{{Review of Particle Physics}}.
\newblock {\it \bibinfo{journal}{Chin. Phys.}\/},  {\it
  \bibinfo{volume}{C38}\/}, \bibinfo{pages}{090001}.
  \DOIprefix\doi{10.1088/1674-1137/38/9/090001}.
\bibitem[{Petropoulou et~al.(2015)Petropoulou, Dimitrakoudis, Padovani,
  Mastichiadis and Resconi}]{Elisa}
\bibinfo{author}{Petropoulou, M.}, \bibinfo{author}{Dimitrakoudis, S.},
  \bibinfo{author}{Padovani, P.}, \bibinfo{author}{Mastichiadis, A.}, and
  \bibinfo{author}{Resconi, E.} (\bibinfo{year}{2015}).
\newblock \bibinfo{title}{{Photohadronic Origin of $\gamma$-ray {BL Lac}
  Emission: Implications for {IceCube} Neutrinos}}.
\newblock {\it \bibinfo{journal}{Monthly Notices of the Royal Astronomical
  Society}\/},  {\it \bibinfo{volume}{448}\/}, \bibinfo{pages}{2412--2429}.
  \DOIprefix\doi{10.1093/mnras/stv179}.
  \href{http://arxiv.org/abs/1501.07115}{\tt arXiv:1501.07115}.
\bibitem[{Prescott et~al.(2015)}]{prescott2015galaxy}
\bibinfo{author}{Prescott, M.} et~al. (\bibinfo{year}{2015}).
\newblock \bibinfo{title}{{Galaxy And Mass Assembly {(GAMA)}: The 325 {MHz}
  Radio Luminosity Function of {AGN} and Star Forming Galaxies}}.
\newblock {\it \bibinfo{journal}{Monthly Notices of the Royal Astronomical
  Society}\/},  {\it \bibinfo{volume}{457}\/}, \bibinfo{pages}{730--744}.
  \DOIprefix\doi{10.1093/mnras/stv3020}.
  \href{http://arxiv.org/abs/1601.00003}{\tt arXiv:1601.00003}.
\bibitem[{Silvestri and Barwick(2010)}]{silvestri2010constraints}
\bibinfo{author}{Silvestri, A.}, and \bibinfo{author}{Barwick, S.~W.}
  (\bibinfo{year}{2010}).
\newblock \bibinfo{title}{{Constraints on Extragalactic Point Source Flux from
  Diffuse Neutrino Limits}}.
\newblock {\it \bibinfo{journal}{Phys. Rev. D}\/},  {\it
  \bibinfo{volume}{81}\/}, \bibinfo{pages}{023001}.
  \DOIprefix\doi{10.1103/PhysRevD.81.023001}.
  \href{http://arxiv.org/abs/0908.4266}{\tt arXiv:0908.4266}.
\bibitem[{Treister et~al.(2009)Treister, Urry and Virani}]{treister}
\bibinfo{author}{Treister, E.}, \bibinfo{author}{Urry, C.~M.}, and
  \bibinfo{author}{Virani, S.} (\bibinfo{year}{2009}).
\newblock \bibinfo{title}{{The Space Density of {Compton}-Thick Active Galactic
  Nucleus and the {X}-Ray Background}}.
\newblock {\it \bibinfo{journal}{Astrophys. J.}\/},  {\it
  \bibinfo{volume}{696}\/}, \bibinfo{pages}{110}.
  \DOIprefix\doi{10.1088/0004-637X/696/1/110}.
  \href{http://arxiv.org/abs/0902.0608}{\tt arXiv:0902.0608}.
\bibitem[{Wright(2006)}]{Wright}
\bibinfo{author}{Wright, E.~L.} (\bibinfo{year}{2006}).
\newblock \bibinfo{title}{{A Cosmology Calculator for the {World} {Wide}
  {Web}}}.
\newblock {\it \bibinfo{journal}{Publications of the Astronomical Society of
  the Pacific}\/},  {\it \bibinfo{volume}{118}\/}, \bibinfo{pages}{1711--1715}.
  \DOIprefix\doi{10.1086/510102}.
  \href{http://arxiv.org/abs/astro-ph/0609593}{\tt arXiv:astro-ph/0609593}.
\bibitem[{Zacharias and Schlickeiser(2012)}]{zacharias2012external}
\bibinfo{author}{Zacharias, M.}, and \bibinfo{author}{Schlickeiser, R.}
  (\bibinfo{year}{2012}).
\newblock \bibinfo{title}{{External Compton Emission in Blazars of Nonlinear
  Synchrotron Self-Compton-Cooled Electrons}}.
\newblock {\it \bibinfo{journal}{Astrophys. J.}\/},  {\it
  \bibinfo{volume}{761}\/}, \bibinfo{pages}{110}.
  \DOIprefix\doi{10.1088/0004-637X/761/2/110}.
  \href{http://arxiv.org/abs/1210.6837}{\tt arXiv:1210.6837}.

\end{thebibliography}

\end{document}